\documentclass[acmsmall, natbib=false, nonacm, screen]{acmart}

\usepackage{graphicx}
\usepackage{tabularx}

\usepackage[T1]{f ontenc}
\usepackage[utf8]{inputenc}

\usepackage{csquotes}
\usepackage{nicefrac}
\usepackage[textsize=tiny]{todonotes}
\usepackage{booktabs}
\usepackage{xcolor}
\usepackage{tikz}
\usetikzlibrary{arrows,positioning}
\usetikzlibrary{shapes.geometric, positioning, quantikz}
\usetikzlibrary{automata, fit}
\usetikzlibrary{positioning}
\usepackage{flushend}
\usepackage{hyperref}
\usepackage[binary-units=true, detect-all=true]{siunitx}
\usepackage{etoolbox}
\usepackage{soul}
\usepackage{amsmath}
\usepackage{bm}
\usepackage{pifont}
\newcommand{\cmark}{\text{\ding{51}}}%
\newcommand{\xmark}{\text{\ding{55}}}%
\robustify\bfseries

\makeatletter
\let\MYcaption\@makecaption
\makeatother
\usepackage[font=footnotesize]{subcaption}
\makeatletter
\let\@makecaption\MYcaption
\makeatother

\usepackage{yquant}
\useyquantlanguage{groups}
\usepackage{pgfplotstable}

\newcommand{\aref}[1]{\hyperref[#1]{Appendix~\ref*{#1}}}

\usepackage[style=ieee, maxnames=3, minnames=1, date=year, url=false, doi=false,isbn=false,backend=biber]{biblatex}
\AtEveryBibitem{%
  \clearname{editor}%
  \clearfield{series}%
  \clearfield{isbn}%
  \clearfield{issn}%
    \clearfield{volume}%
  \clearfield{number}%
  \clearfield{pages}%
}

\DeclareFieldFormat
  [misc, book]
  {title}{\mkbibquote{#1}}

\newtheorem{example}{Example}

\authorsaddresses{}

\definecolor{color1}{RGB}{33, 145, 140}
\definecolor{color2}{RGB}{64,1,84}

\author{Nils Quetschlich}
\orcid{0000-0002-4369-5207}
\affiliation{
  \department{Chair for Design Automation}
  \institution{Technical University of Munich}
  \city{Munich}
  \country{Germany}
}
\email{nils.quetschlich@tum.de}
\author{Lukas Burgholzer}
\orcid{0000-0003-4699-1316}
\affiliation{%
  \department{Chair for Design Automation}
  \institution{Technical University of Munich}
  \city{Munich}
  \country{Germany}
}
\email{lukas.burgholzer@tum.de}
\author{Robert Wille}
\orcid{0000-0002-4993-7860}
\affiliation{%
  \institution{Technical University of Munich}
  \department{Chair for Design Automation}
  \city{Munich}
  \country{Germany}
}
\affiliation{%
  \institution{Software Competence Center Hagenberg GmbH}
  \city{Hagenberg}
  \country{Austria}
}
\email{robert.wille@tum.de}
\begin{document}

\title{MQT~Predictor: Automatic Device Selection with Device-Specific Circuit Compilation for Quantum Computing}

\begin{abstract}
Fueled by recent accomplishments in quantum computing hardware and software, an increasing number of problems from various application domains are being explored as potential use cases for this new technology.
Similarly to classical computing, realizing an application on a particular quantum device requires the corresponding (quantum) circuit to be \emph{compiled} so that it can be executed on the device.
With a steadily growing number of available devices---each with their own advantages and disadvantages---and a wide variety of different compilation tools, the number of choices to consider when trying to realize an application is quickly exploding.
Due to missing tool support and automation, especially \mbox{end-users} who are not quantum computing experts are easily left unsupported and overwhelmed.

In this work, we propose a methodology that allows one to \emph{automatically select} a suitable quantum device for a particular application \emph{and} provides an \emph{optimized compiler} for the selected device.
The resulting framework---called the \emph{MQT~Predictor}---not only supports \mbox{end-users} in navigating the vast landscape of choices, it also allows \mbox{\emph{mixing and matching}} compiler passes from various tools to create optimized compilers that transcend the individual tools.
Evaluations of an exemplary framework instantiation based on more than $500$ quantum circuits and seven devices have shown that---compared to both Qiskit's and TKET's most optimized compilation flows for all devices---the MQT~Predictor produces circuits within the \mbox{top-3} out of $14$ baselines in more than $98\%$ of cases while frequently outperforming any tested combination by up to $53\%$ when optimizing for \emph{expected fidelity}.
Additionally, the framework is trained and evaluated for \emph{critical depth} as another \emph{figure of merit} to showcase its flexibility and generalizability---producing circuits within the \mbox{top-3} in $89\%$ of cases while frequently outperforming any tested combination by up to $400\%$.
MQT~Predictor is part of the \emph{Munich Quantum Toolkit}~(MQT) and publicly available as \mbox{open-source} on GitHub (\url{https://github.com/cda-tum/mqt-predictor}) and as an \mbox{easy-to-use} \emph{Python} package (\url{https://pypi.org/p/mqt.predictor}).
\end{abstract}

\maketitle

\section{Introduction}\label{sec:intro}
Quantum computing is an emerging computational technology and has seen considerable accomplishments in both hardware and software---especially in recent years.
New quantum devices with an increasing number of qubits and higher fidelity rates are emerging almost on a daily basis.
A similar trend can be observed for the development of quantum \emph{Software Development Kits}~(SDKs) such as, e.g., IBM's Qiskit~\cite{qiskit}, Quantinuum's TKET~\cite{sivarajahKetRetargetableCompiler2020}, Google's Cirq~\cite{cirq}, Xanadu's PennyLane~\cite{bergholmPennyLaneAutomaticDifferentiation2020}, or Rigetti's Forest~\cite{rigetti}.
This sparks interest not only in academia but also in industry---with promising applications of quantum computing emerging in various domains, such as finance~\cite{stamatopoulosOptionPricingUsing2020}, optimization~\cite{quetschlich2023satellite}, machine learning~\cite{zoufalQuantumGenerativeAdversarial2019}, or chemistry~\cite{peruzzoVariationalEigenvalueSolver2014}.
However, realizing such quantum computing-based solutions remains a challenge that is not \mbox{straight-forward} to solve and, to date, requires several manual steps~\cite{quetschlichAutomatedFrameworkRealizing2023}.

First, a suitable quantum device must be \emph{selected} for the execution of the developed quantum algorithm.
This alone is \mbox{non-trivial} since new quantum devices based on various underlying technologies emerge almost daily---each with their own advantages and disadvantages.
There are hardly any practical guidelines on which device to choose based on the targeted application.
As such, the best guess in many cases today is to simply try out many (if not all) possible devices and, afterwards, choose the best results---certainly a time- and resource-consuming endeavor that is not sustainable for the future.

Once a target device has been selected, the quantum circuit, which is typically designed in a \mbox{device-agnostic} fashion that does not account for any hardware limitations (such as a limited \mbox{gate-set} or limited connectivity), must be \emph{compiled} accordingly so that it actually becomes executable on that device.
This is similar to classical computing, where high-level, platform-agnostic code (e.g., written in C++) is being \emph{compiled} to low-level, platform-specific machine code before being executed on a particular classical computer.
Compilation itself is a sequential process consisting of a sequence of \emph{compilation passes} that, \mbox{step-by-step}, transform the original quantum circuit so that it eventually conforms to the limitations imposed by the target device.
Since many of the underlying problems in compilation are computationally hard (e.g., satisfying the connectivity limitations of a certain device with the least amount of overhead being NP-complete~\cite{siraichiQubitAllocation2018}), there is an ever-growing variety of compilation passes available across several quantum SDKs and software tools---again, each with their own advantages and disadvantages.
As a result of the sheer number of options, choosing the best sequence of compilation passes for a given application is nearly impossible.
Consequently, most quantum SDKs (such as Qiskit and TKET) provide \mbox{easy-to-use} high-level function calls that encapsulate \enquote{their} sequence of compilation passes into a single compilation flow.
While this allows to conveniently compile circuits, it has several drawbacks.
For one, it creates a kind of \emph{vendor lock} that limits the available compilation passes to those available in the SDK offering the compilation flow.
Furthermore, the respective compilation flows are designed to be broadly applicable and, hence, are neither device-specific nor circuit-specific.
On top of that, no means are provided to optimize for a specific \emph{figure of merit}---only the degree of desired optimization can be specified without any explicit optimization criterion.

Overall, this leaves anyone trying to realize a quantum computing application with more options than they could ever feasibly explore.
What makes things worse is that the choice of the quantum device and the subsequent compilation flow heavily affect the quality of the resulting circuit and its execution.
Choosing the right combination can often make the difference between a useful result and pure noise.
This is especially unfortunate for \mbox{end-users} who are not experts in quantum computing and just want to use the technology to solve their application domain problem.

In this work\footnote{Preliminary versions of this work have been published in \cite{quetschlich2023prediction, quetschlich2023compileroptimization}.}, a new approach to \emph{automate} the quantum device selection and according compilation is proposed to free \mbox{end-users} from this task and to prevent a scenario where quantum computers can only be utilized by quantum computing experts---hindering the overall adoption of that technology.
To this end, we tackle this problem from two angles:
\begin{enumerate}
	\item We propose a \emph{prediction} method (based on \emph{Supervised Machine Learning}) that, without performing any compilation, automatically predicts the most suitable device for a given application. This completely eliminates the manual and laborious task of determining a suitable target device and guides \mbox{end-users} through the vast landscape of choices without the need for quantum computing expertise.
	\item We propose a method (based on \emph{Reinforcement Learning}) that produces device-specific quantum circuit compilers by combining compilation passes from various compiler tools and learning optimized sequences of those passes with respect to a customizable \emph{figure of merit}.
	This \mbox{\emph{mix-and-match}} of compiler passes from various tools allows one to eliminate vendor locks and to create optimized compilers that transcend the individual tools.
\end{enumerate}
Combining both of these methods into a holistic framework---the \emph{MQT~Predictor}---allows one to \emph{automatically select} a suitable quantum device for a particular application while also providing an \emph{optimized compiler} for the selected device and a customizable figure of merit.

\vspace{2cm}

To show its benefits, an instantiation of the proposed approach is implemented based on more than $500$ quantum circuits ranging from $2$ to $90$ qubits and considering seven different quantum devices based on two qubit technologies.
As a baseline to compare against, the most optimized compilation flows available in Qiskit and TKET have been used---leading to a total of $7\times 2=14$ possible combinations of device and compiler.
Compared to these reference results, the MQT~Predictor produces circuits with an expected fidelity that is \emph{on par with the best} possible result that could be achieved by trying out all combinations of devices and compilers.
Specifically, the \emph{automatic device selection} and \emph{optimized circuit compilation} produced circuits within the \mbox{top-3} in more than $98\%$ of cases while frequently outperforming any tested combination by up to $53\%$.
Additionally, MQT~Predictor is trained and evaluated for \emph{critical depth} as another figure of merit to showcase its flexibility and generalizability.
The respective results show that the proposed method also achieves a similar performance---producing circuits within the \mbox{top-3} in $89\%$ of cases while frequently outperforming any tested combination by up to $400\%$.
The corresponding framework (which is part of the \emph{Munich Quantum Toolkit}~(MQT)~\cite{willeMQTHandbookSummary2024}) including the \mbox{pre-trained} models and classifiers is publicly available as \mbox{open-source} on GitHub (\url{https://github.com/cda-tum/mqt-predictor}) and as an \mbox{easy-to-use} \emph{Python} package (\url{https://pypi.org/p/mqt.predictor}).

The rest of this work is structured as follows: In \autoref{sec:background}, quantum circuit compilation flows and related work are reviewed.
Based on that, \autoref{sec:general_ideas} introduces the proposed approach and the underlying methodologies with details given in \autoref{sec:rl} for the compilation using reinforcement learning and \autoref{sec:ml} for the device selection using supervised machine learning.
The resulting combined approach is then described in \autoref{sec:mqtpredictor}, evaluated in \autoref{sec:evaluation}, and discussed in \autoref{sec:discussion}.
\autoref{sec:conclusions} concludes this work.

\section{Background and Motivation}\label{sec:background}
To keep this work self-contained, this section gives a brief introduction to quantum circuit compilation, discusses its challenges, and reviews the related work.

\subsection{Quantum Circuit Compilation}\label{sec:compilation}

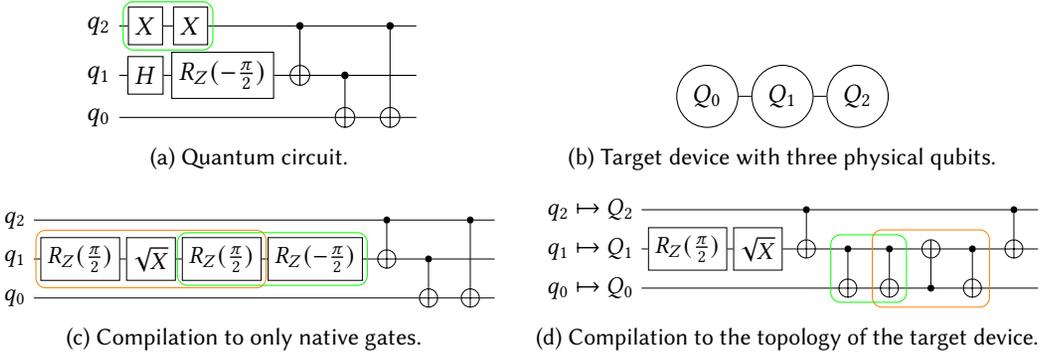
\begin{figure}[t]
     \centering
     \begin{subfigure}{0.49\linewidth}
     \centering
				\begin{tikzpicture}
				  \begin{yquant}
				    	qubit {${q_2}$} q;
						qubit {${q_1}$} q[+1];
						qubit {${q_0}$} q[+1];

						[name=left]
				    	box {$X$} q[0];
				    	[name=right]
				    	box {$X$} q[0];

				    	box {$H$} q[1];
				    	box {$R_Z(-\frac{\pi}{2})$} q[1];

				    	cnot q[1] | q[0];
				    	cnot q[2] | q[1];
				    	cnot q[2] | q[0];

				  \end{yquant}
				  \node[fit=(left)(right), draw, green, rounded corners, inner sep=2pt] {};
				\end{tikzpicture}
		\caption{Quantum circuit.}
         \label{fig:sub_1}
		 \end{subfigure}
		 \hfill
		       \begin{subfigure}{0.49\linewidth}
     \centering
					   \begin{tikzpicture}
						    \node (p0) at ( 0, 1) [circle, draw]{$Q_0$};
						    \node (p1) at ( 1,1) [circle, draw]{$Q_1$};
						    \node (p2) at ( 2,1) [circle, draw]{$Q_2$};
						    \begin{scope}[every path/.style={-}]
						       \draw (p0) -- (p1);
						       \draw (p1) -- (p2);
						    \end{scope}
						\end{tikzpicture}

			         \caption{Target device with three physical qubits.}
			         \label{fig:arch}
		     \end{subfigure}

		     \vspace{3mm}
\begin{subfigure}{0.48\linewidth}
     \centering
\resizebox{1.0\linewidth}{!}{
				\begin{tikzpicture}
				  \begin{yquant}
				    	qubit {${q_2}$} q;
						qubit {${q_1}$} q[+1];
						qubit {${q_0}$} q[+1];

						[name=left_decomp]
				    	box {$R_Z(\frac{\pi}{2})$} q[1];
				    	box {$\sqrt{X}$} q[1];
						[name=left]
				    	box {$R_Z(\frac{\pi}{2})$} q[1];
				    	[name=right]
				    	box {$R_Z(-\frac{\pi}{2})$} q[1];

				    	cnot q[1] | q[0];
				    	cnot q[2] | q[1];
				    	cnot q[2] | q[0];

				  \end{yquant}
				  \node[fit=(left)(right), draw, green, rounded corners, inner sep=2pt] {};
				  \node[fit=(left_decomp)(left), draw, orange, inner ysep=3pt, inner xsep=2pt, rounded corners] {};
				\end{tikzpicture}}
         \caption{Compilation to only native gates.}
         \label{fig:sub_2}
     \end{subfigure}
		     \hfill
     \begin{subfigure}{0.48\linewidth}
     \centering
\resizebox{1.00\linewidth}{!}{
		\begin{tikzpicture}
  \begin{yquant}
    	qubit {${q_2} \mapsto Q_2$} q;
		qubit {${q_1} \mapsto Q_1$} q[+1];
		qubit {${q_0} \mapsto Q_0$} q[+1];

		box {$R_Z(\frac{\pi}{2})$} q[1];
		box {$\sqrt{X}$} q[1];

	    cnot q[1] | q[0];
		[name=left]

		cnot q[2] | q[1];
		[name=right]
		cnot q[2] | q[1];
		cnot q[1] | q[2];
		[name=right_decomp]
		cnot q[2] | q[1];
		cnot q[1] | q[0];

  \end{yquant}
\node[fit=(left)(right), draw, green, inner ysep=10pt, yshift=0.28cm, rounded corners] {};
				  \node[fit=(right)(right_decomp), draw, orange, yshift=0.28cm, inner ysep=12pt, rounded corners] {};
\end{tikzpicture}}
         \caption{Compilation to the topology of the target device.}
         \label{fig:sub_3}
     \end{subfigure}
        \caption{Compilation of a quantum circuit to a targeted device.}
        \label{fig:compflow}
\end{figure}

Solving any problem using quantum computing requires encoding the problem as a quantum algorithm.
These algorithms are typically described as sequences of quantum operations (often referred to as \emph{quantum gates}) that form a \emph{quantum circuit}.
Usually, one distinguishes between \emph{single-qubit} gates affecting only one qubit and \emph{multi-qubit} gates affecting more than one qubit at the same time.

\begin{example}\label{ex:compflow1}
\autoref{fig:sub_1} shows a small exemplary quantum circuit with three qubits ($q_0$, $q_1$, $q_2$) and four different kinds of quantum gates acting on them.
There are two \mbox{single-qubit} gates on $q_2$ (two \emph{NOT} gates denoted as \emph{X}) and two \mbox{single-qubit} gates on $q_1$ (a Hadamard gate and a $Z$-rotation gate denoted as \emph{H} and \emph{$R_Z$}, respectively).
Additionally, there are \mbox{multi-qubit} gates between all possible qubit pairs, namely \emph{controlled-NOT} (CNOT) gates denoted as $\bullet$ and $\oplus$ for their control, respectively, target qubit.
\end{example}

The actual compilation can only be started as soon as a suitable quantum device is \emph{selected} and, by that, its native \mbox{gate-set} and connectivity constraints are known.
The device selection itself is already challenging due to several factors:

\begin{itemize}
	\item Various underlying quantum technologies are currently pursued in parallel with no clear winner yet and all of them have their own advantages and disadvantages. \emph{Ion trap}-based devices, e.g., feature an all-to-all connectivity but provide rather slow gate execution times while, e.g., \emph{superconducting}-based devices provide fast gate execution times but usually come with a restricted connectivity.
	\item Even devices based on the same technology significantly vary in their hardware characteristics, such as qubit count, their respective error rates, coherence times, qubit connectivity, and gate execution times.
	\item On top of all that, the domain of quantum computing is fast-paced and constantly changing---quickly and frequently redefining the state of the art.
\end{itemize}

\begin{example}\label{ex:compflow1_5}
\autoref{fig:arch} shows the topology of a \mbox{three-qubit} quantum device. This topology indicates that \mbox{two-qubit} gates may only be applied to $(Q_0, Q_1)$ or $(Q_1, Q_2)$, but not to $(Q_0, Q_2)$.
\end{example}

As introduced in \autoref{sec:intro}, executing such quantum circuits on an actual quantum device requires compilation, which is usually conducted by \emph{compilers} that are part of quantum SDKs as \mbox{easy-to-use} \mbox{high-level} function calls.
However, compilation actually is a sequential process that can be roughly divided into
three kinds of \emph{compilation passes}: \emph{Synthesis} passes compile a given quantum circuit to the native \mbox{gate-set} of a selected quantum device, \emph{mapping} passes enforce its connectivity constraints, and \emph{optimization} passes exploit various possibilities to reduce the circuit's complexity.
Quantum SDKs usually offer a wide range of \mbox{device-independent} optimization passes with the goal of creating a more efficient quantum circuit which still realizes the same functionality by, e.g., \mbox{gate-cancellation}, \mbox{gate-commutation}, or \mbox{high-level} synthesis routines.

\begin{example}\label{ex:compflow2}
	Since $X\cdot X = I$, the circuit shown in \autoref{fig:sub_1} can be simplified by canceling the two subsequent \emph{X} gates (denoted by the green box).
\end{example}

Usually, those device-independent optimization passes are followed by the \emph{device-dependent} compilation passes.
Every quantum device can only directly execute a subset of all possible quantum gates---its so-called native gates.
These native gates usually comprise a handful of \mbox{single-qubit} gates and one \mbox{multi-qubit} gate to become a \emph{universal} \mbox{gate-set}.
Therefore, all non-native gates must be \emph{synthesized} into a sequence of native gates of that respective quantum device---a procedure that is \mbox{NP-complete} to solve in an optimal fashion~\cite{pehamDepthoptimalSynthesisClifford2023}.
Consequently, this alters the quantum circuit---leading to further room for optimization.

\begin{example}\label{ex:compflow3}
Assume that the targeted device from \autoref{ex:compflow1_5} has a native \mbox{gate-set} consisting of \emph{$R_Z$}, \emph{$\sqrt{X}$}, \emph{X}, and \emph{CNOT} gates and that the circuit from \autoref{fig:sub_1} shall be executed on it.
Only the Hadamard gate on qubit $q_1$ is not a native gate of the device and, hence, must be synthesized into a sequence of such gates.
The orange box in \autoref{fig:sub_2} indicates one such decomposition into two $R_Z(\frac{\pi}{2})$ and a $\sqrt{X}$ gate.
Again, this circuit alteration leads to further optimization potential since the two subsequent $R_Z$ gates in the green box, again, cancel each other out.
\end{example}

In addition to the limited \mbox{gate-set}, many quantum devices (e.g., based on such as superconducting qubits) only have a \emph{limited connectivity} between their underlying qubits---requiring that all \mbox{multi-qubit} gates must be placed on qubits which are actually connected on the respective device.
To adapt a quantum circuit with gates operating between arbitrary qubits to such a limited connectivity, a \emph{mapping} pass is necessary.
This pass is often split into \emph{layouting} and \emph{routing}.
The former assigns each (logical) qubit of the circuit to a (physical) qubit on the device and the latter ensures that all connectivity constraints are satisfied by changing the qubit assignment throughout the circuit, e.g., using \emph{SWAP} gates---a procedure that is, again, \mbox{NP-complete} to solve in an optimal fashion~\cite{siraichiQubitAllocation2018}.
Then, again, optimization passes can be applied to reduce the complexity of the resulting quantum circuit.

\begin{example}\label{ex:compflow4}
Consider again the circuit shown in \autoref{fig:sub_2} which shall be executed on the \mbox{three-qubit} quantum device shown in \autoref{fig:arch}.
Since the quantum circuit contains \mbox{multi-qubit} gates between all qubit pairs, there is no trivial layout that satisfies the connectivity constraints throughout the entire circuit.
Therefore, the circuit must be mapped by introducing a \emph{SWAP} gate as denoted in the orange box in \autoref{fig:sub_3} (again, compiled into its native \mbox{gate-set} sequence of three \emph{CNOT} gates).
Although the resulting quantum circuit is now fully executable, a successive optimization pass can, again, eliminate two subsequent \emph{CNOT} gates.
\end{example}

Although the compilation process is rather \mbox{straight-forward} conceptually---consisting of \emph{synthesis}, \emph{mapping}, and \emph{optimization} passes---it comes with practical challenges.
For each step, a variety of different techniques and approaches have been proposed, such as~\cite{gilesExactSynthesisMultiqubit2013, amyMeetinthemiddleAlgorithmFast2013, millerElementaryQuantumGate2011, zulehnerOnepassDesignReversible2018, niemann2020advancedexactsynt, pehamDepthoptimalSynthesisClifford2023} for synthesis, ~\cite{zulehnerEfficientMethodologyMapping2019,
matsuoEfficientMethodQuantum2019,
willeMappingQuantumCircuits2019,
liTacklingQubitMapping2019,
pehamOptimalSubarchitecturesQuantum2023,
hillmichExploitingQuantumTeleportation2021,
zulehnerCompilingSUQuantum2019, willeMQTQMAPEfficient2023, schmidHybridCircuitMapping2024} for mapping, and ~\cite{hattoriQuantumCircuitOptimization2018, vidalUniversalQuantumCircuit2004, itokoQuantumCircuitCompilers2019, maslovQuantumCircuitSimplification2008} for optimization.
Therefore, \mbox{end-users} are easily overwhelmed with selecting the \emph{best} combination of provided SDKs and respective compilation passes---not even mentioning the effort of becoming acquainted with multiple (often poorly documented) SDKs and, respectively, needed conversions between them.
Every SDK usually implements a subset of these compilation passes with varying degrees of effectiveness and efficiency.
Those passes are provided as \mbox{pre-configured}, fixed sequences of compilation passes (for various levels of desired optimization)---independent of the given quantum circuit.

\subsection{Related Work}\label{sec:related}
A handful of techniques have already been explored to support the quantum device selection and compilation.
In~\mbox{\cite{salm_automating_2021, kharkov2022arline, millsApplicationMotivatedHolisticBenchmarking2021, lubinskiApplicationOrientedPerformanceBenchmarks2021a}},
approaches that compile a given circuit for \emph{all} devices in a \mbox{brute-force} fashion and compare their resulting quality (also referred to as \emph{autotuning}) are proposed.
A similar approach to select the best sequence of compilation passes by evaluating all possible combinations has been proposed in~\cite{dangwal2023clifford}.
Furthermore, in~\cite{salm_nisq_2020}, a methodology has been proposed to determine whether compiled circuits can be reliably executed on certain devices to give some guidance.
The same authors have further explored how \emph{\mbox{Multi-Criteria} Decision Analysis} methods can be used to select the most suitable compiled circuit from a given set of all compiled circuits~\cite{salmPrio2022} and how machine learning can be used to predict the goodness of a device and compiler combination~\cite{salmcloser23}.

\emph{Supervised Machine Learning}~(ML) techniques have also been applied to quantum compilation, e.g., in~\cite{ml_alexandru}, an ML-based layout and routing scheme is proposed, while
in~\cite{mete2022ml_opt}, an ML model is trained to predict the optimization potential of a given quantum circuit to decide whether to apply (potentially costly) optimization passes or not.
Furthermore, \emph{Reinforcement Learning} (RL) has been explored in this context, e.g., in~\cite{rloptgoogle2021} where a strategy for applying individual gate transformation rules to optimize quantum circuits is learned or in~\cite{wangQuEestGraphTransformer2022} where an \mbox{RL model} is trained to learn how to estimate the circuit fidelity for specific devices.
Moreover, environments to train \mbox{RL models} for various compilation tasks are proposed in~\cite{vanderlinde2023qgym}---similar to the environment proposed in~\cite{quetschlich2023compileroptimization}.

\vspace{1cm}

In classical compilation, machine learning has been explored for a much longer time which led to sophisticated compiler optimization techniques with overviews given in, e.g., \cite{ml_in_compilers, autotuning_compiler} and established these methods as \mbox{state-of-the-art} techniques in this domain.
Additionally, even a combination of these techniques has been explored in~\cite{agakov_comb} by using a machine learning model to determine promising areas of optimization space which then can be further investigated using autotuning.
On top of that, reinforcement learning approaches targeting the~LLVM~\cite{llvm} compiler emerged especially in recent years with encouraging results~\cite{reinforcement1, reinforcement2, reinforcement3}.

So far, many of the approaches toward both quantum device selection and compilation~\cite{salm_automating_2021, kharkov2022arline, salm_nisq_2020, millsApplicationMotivatedHolisticBenchmarking2021, lubinskiApplicationOrientedPerformanceBenchmarks2021a, salmPrio2022, dangwal2023clifford} follow a naive \mbox{brute-force} approach that, obviously, is not a \mbox{long-term} solution due to the steadily increasing number of quantum devices and compilers.
Furthermore, existing ML-based tools to support \mbox{end-users}~\cite{salmcloser23, ml_alexandru, mete2022ml_opt, rloptgoogle2021, wangQuEestGraphTransformer2022} do not consider the entire compilation or do not provide a broad coverage in terms of supported devices, range of qubits, or considered algorithms, and frequently lack the necessary means to scale to significantly more devices and compilers.
On top of that, most approaches focus on using current compilers \emph{\mbox{as-is}} and hardly make use of the sophisticated \mbox{ML-based} approaches for classical compiler optimization---leading to lots of untapped potential.

\section{General Idea}\label{sec:general_ideas}
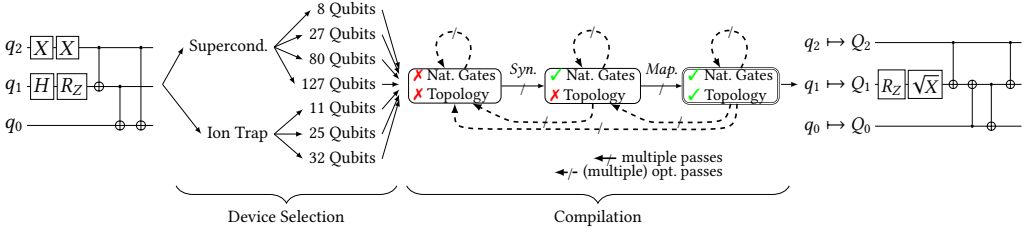
\begin{figure}
\centering
\resizebox{1.0\linewidth}{!}{
\begin{tikzpicture}
  \node (start) at (-6.0,-0.5)   {
\resizebox{0.23\linewidth}{!}{
	\begin{tikzpicture}
				  \begin{yquant}[register/minimum height=1cm]
				    	qubit {\huge{${q_2}$}} q;
						qubit {\huge{${q_1}$}} q[+1];
						qubit {\huge{${q_0}$}} q[+1];

				    	box {\huge{$X$}} q[0];
				    	box {\huge{$X$}} q[0];

				    	box {\huge{$H$}\vphantom{$R_Z$}} q[1];
				    	box {\huge{$R_Z$}} q[1];

				    	cnot q[1] | q[0];
				    	cnot q[2] | q[1];
				    	cnot q[2] | q[0];

				  \end{yquant}
				\end{tikzpicture} }
  };

\def\posidevs{-2.0}
  \node[anchor=east] (superc) at (\posidevs,0.25)   {Supercond.};
  \node[anchor=east] (ion) at (\posidevs,-1.5)   {Ion Trap};

\def\posi{0.2}
  \node[anchor=east] (8) at (\posi,1)   {8 Qubits};
  \node[anchor=east] (27) at (\posi,0.5)   {27 Qubits};
  \node[anchor=east] (80) at (\posi,0)   {80 Qubits};
  \node[anchor=east] (127) at (\posi,-0.5)   {127 Qubits};
  \node[anchor=east] (11) at (\posi,-1.0)   {11 Qubits};
  \node[anchor=east] (25) at (\posi,-1.5)   {25 Qubits};
  \node[anchor=east] (32) at (\posi,-2.0)   {32 Qubits};

  \node[anchor=east] (rl) at (0.8,-0.5)   {};

  \draw[-latex] (start.east) -- (ion.west);
  \draw[-latex] (start.east) -- (superc.west);
  \draw[-latex] (ion.east) -- (11.west);
  \draw[-latex] (ion.east) -- (25.west);
  \draw[-latex] (ion.east) -- (32.west);
  \draw[-latex] (superc.east) -- (8.west);
  \draw[-latex] (superc.east) -- (27.west);
  \draw[-latex] (superc.east) -- (80.west);
  \draw[-latex] (superc.east) -- (127.west);

  \draw[-latex] (8.east) -- ([xshift=0mm, yshift=3mm]rl.west);
  \draw[-latex] (27.east) -- ([xshift=0mm, yshift=2mm]rl.west);
  \draw[-latex] (80.east) -- ([xshift=0mm, yshift=1mm]rl.west);
  \draw[-latex] (127.east) -- ([xshift=0mm, yshift=0mm]rl.west);
  \draw[-latex] (11.east) -- ([xshift=0mm, yshift=-1mm]rl.west);
  \draw[-latex] (25.east) -- ([xshift=0mm, yshift=-2mm]rl.west);
  \draw[-latex] (32.east) -- ([xshift=0mm, yshift=-3mm]rl.west);

\node[right of = rl, xshift=2.8cm, inner ysep=0mm, yshift=-3mm] (rl_mdp){
\resizebox{0.55\linewidth}{!}{
\begin{tikzpicture}
\def\w2{1.0pt}
\def\minsize{0.6cm}
\def\height_mdp2{-5}
    \def\w2{1.0pt}
\def\minsize{0.6cm}

	\node (p0) at ( 0, \height_mdp2) [minimum size=\minsize,rectangle,rounded corners,draw,align=left]{\textcolor{red}{$\bm{\xmark}$} Nat. Gates\\\textcolor{red}{$\bm{\xmark}$} Topology};
    \node (p3) at ( 3,\height_mdp2)  [minimum size=\minsize,rectangle,rounded corners,draw,align=left]{\textcolor{green}{$\bm{\cmark}$} Nat. Gates\\\textcolor{red}{$\bm{\xmark}$} Topology};
    \node (p4) at ( 6,\height_mdp2) [minimum size=\minsize,rectangle,rounded corners, double,draw,align=left]{\textcolor{green}{$\bm{\cmark}$} Nat. Gates\\\textcolor{green}{$\bm{\cmark}$} Topology};

    \path[-latex] (p0) edge [] node [align=center, above, yshift=-1.9mm]  {\emph{Syn.} \\ /} (p3);
    \path[-latex] (p3) edge [] node [align=center, above, yshift=-1.9mm]  {\emph{Map.} \\ /} (p4);

    \def\lin{120}
    \def\lout{60}
    \def\lloose{7}
    \tikzset{every loop/.style={in=\lin,out=\lout,looseness=\lloose}}

    \path[-latex]  (p0) edge [dashed, loop above, line width=\w2] node [align=center, above, yshift=-1.5mm] {/}(p0);
    \path[-latex]  (p3) edge [dashed, loop above, line width=\w2] node [align=center, above, yshift=-1.5mm]{/}(p3);
    \path[-latex]  (p4) edge [dashed, loop above, line width=\w2] node [align=center, above, yshift=-1.5mm]{/}(p4);

\def\h{0.4}
\draw[-latex] (p4.270) [dashed, line width=\w2] .. controls ($(p4.south) - (0.1,\h)$) .. node [align=center, above, yshift=-1.9mm, xshift=-9mm]{/} ($(p4.south) - (1, \h)$) -- ($(p3.south) - (-1.5, \h)$) .. controls ($(p3.south) - (-1.2, \h)$) .. (p3.315);
\draw[-latex] (p3.270) [dashed, line width=\w2] .. controls ($(p3.south) - (0,\h)$) .. node [align=center, above, yshift=-1.9mm, xshift=-9mm]{/} ($(p3.south) - (1, \h)$) -- ($(p0.south) - (-1.5, \h)$) .. controls ($(p0.south) - (-1.2, \h)$) .. (p0.315);
\def\hd{0.55}
\draw[-latex] (p4.290) [dashed, line width=\w2] .. controls ($(p4.south) - (-0.1,\hd)$) .. node [align=center, above, yshift=-1.9mm, xshift=-28mm]{/} ($(p4.south) - (1, \hd)$) -- ($(p0.south) - (-1.2, \hd)$) .. controls ($(p0.south) - (0, \hd)$) .. (p0.270);
\node (desc) [text=black, below left=1.0cm and -1cm  of  p4] {multiple passes};
\path[-latex, line width=\w2] (desc.west) edge [] node [align=center, above, yshift=-1.8mm, xshift=1mm] {/} ($(desc.west) - (0.5, 0.0)$);
\node (desc2) [text=black, below left=1.3cm and -1cm  of  p4] {(multiple) opt. passes};
\path[-latex, line width=\w2] (desc2.west) edge [dashed] node [align=center, above, yshift=-1.8mm, xshift=1mm] {/} ($(desc2.west) - (0.5, 0.0)$);

\end{tikzpicture}}
  };
\draw [decorate,decoration={brace,amplitude=10pt,mirror,raise=4pt},yshift=-2mm]
(-4.0,-2.2) -- (0.5,-2.2) node [black,midway,below, yshift=-5mm] {Device Selection};
\draw [decorate,decoration={brace,amplitude=10pt,mirror,raise=4pt},yshift=-2mm]
(0.7,-2.2) -- (8.5,-2.2) node [black,midway,below, yshift=-5mm] {Compilation};
)
  \node (compiled) [right of = start, xshift=16.0cm]{
\resizebox{0.33\linewidth}{!}{
	\begin{tikzpicture}
				  \begin{yquant}[register/minimum height=1.2cm]
            qubit {\huge{${q_2} \mapsto Q_2$}} q;
		qubit {\huge{${q_1} \mapsto Q_1$}} q[+1];
		qubit {\huge{${q_0} \mapsto Q_0$}} q[+1];
		box {\huge{$R_Z$\vphantom{$\sqrt{X}$}}} q[1];
		box {\huge{$\sqrt{X}$}} q[1];

	    cnot q[1] | q[0];
		cnot q[1] | q[2];
		cnot q[2] | q[1];
		cnot q[1] | q[0];

				  \end{yquant}
				\end{tikzpicture} }
  };

  \draw[-latex]($(compiled.west)-(0.25,0)$) -- ($(compiled.west)+(0.1,0)$);
\end{tikzpicture}

}
	\caption{The device selection and compilation process.}
	\label{fig:search_space}
\end{figure}

In this work, we propose a new approach to \emph{automate} the device selection and compilation process that learns from previously conducted compilations and \emph{predicts} the most suitable target device for a given quantum circuit \emph{as well as} provides an \emph{optimized compiler} for the selected device.
To this end, this section motivates the proposed approach and describes its general ideas, the chosen methodology, and its utilization.

\subsection{Proposed Approach}\label{sec:proposed}

In an ideal scenario, a quantum circuit would be compiled for all possible sequences of compilations on all possible devices to determine the best compilation flow.
However, since compilation comes with considerable complexity both in time and resources, this is practically infeasible.

\begin{example}\label{ex:idea_1}
Assume that we want to find the best compiled version of the quantum circuit shown in \autoref{fig:sub_1} given access to seven different devices (based on two qubit technologies) and various compiler passes provided by multiple SDKs as illustrated in \autoref{fig:search_space}.
Manually determining the best possible combination of options is not only challenging, but close to impossible due to the number of possible combinations---which, in theory, is infinitely large due to backward loops in the compilation process that are introduced by optimization passes.
\mbox{End-users} could just choose \emph{any} (random) path from the \mbox{left-hand} side of \autoref{fig:search_space} to the \mbox{right-hand} side.
However, this is highly unlikely to yield a sufficiently good result (not to dare, the best possible result).
\end{example}

In the proposed approach, we first consider finding the best sequence of \emph{compilation} passes for a particular device as an isolated problem that is independent of the device selection task itself.
To this end, compilation is modeled as a sequential (\emph{Markov Decision)} process of synthesis, mapping, and optimization passes for a given native \mbox{gate-set} and qubit connectivity.
Using \emph{Reinforcement Learning}~(RL), respective models can be trained to \emph{learn} the sequences that lead to the most promising compilation result for each supported device
according to some \emph{figure of merit} such as expected fidelity or critical depth.
These models can then be used as optimized \mbox{black-box} compilers for the respective devices.

Based on these optimized compilers, the task of choosing the most promising device is tackled.
To this end,
\emph{Supervised Machine Learning} (ML) is applied by training an agent that \emph{learns} the best suited device for a given circuit assuming that it will be compiled using the trained RL models.
Thus, a holistic approach is created that takes a \mbox{to-be-compiled} quantum circuit as input and predicts which quantum device to use, compiles it accordingly using the trained RL model, and returns the compiled circuit with detailed compilation information.
All of this is conducted in a fully automated fashion which is especially helpful for end-users who are not experts in quantum computing.
Additionally, compilation passes from various sources can be integrated---effectively eliminating vendor locks by mixing and matching compiler passes from various SDKs.

To learn how to \emph{best} compile quantum circuits, it is necessary to give the notion of \enquote{best} a meaning by defining a \emph{figure of merit} to optimize for.
As this is an essential part of the proposed approach, more details on this way of customizing the compilation are provided in the following.

\subsection{Figures of Merit}\label{sec:eval_metric}
In general, the figure of merit can be completely customizable and consider \emph{anything} from (compiled) circuit characteristics such as gate counts to \emph{soft} factors---applicable to the device selection---such as actual execution costs or the queue length for a specific device vendor.
Nevertheless, a comprehensive figure of merit should consider at least the characteristics of the compiled quantum circuit and the characteristics of the selected quantum device.
Therefore, the respective hardware information needs to be collected, such as, e.g., gate/readout fidelities, gate execution times, or decoherence times.
While for some devices, this information may be publicly available, for other devices, it may be estimated from comparable devices, previous records, or insider knowledge.

\begin{example}
Consider the figure of merit that takes into account the following two aspects:
\begin{enumerate}
\item If the selected device is not large enough to execute a given quantum circuit with respective to its number of qubits, the \mbox{worst-possible} score is assigned.
\item If the device is large enough, the evaluation score is calculated using the formula:

\[
\mathit{\mathcal{F}}=\prod_{i=1}^{|G|} \mathit{\mathcal{F}}(g_i) \prod_{j=1}^{m} \mathit{\mathcal{F}_{RO}}(q_j)
\]
with $\mathit{\mathcal{F}}(g_i)$ being the expected execution fidelity of gate~$g_i$ on its corresponding qubit(s),
$\mathit{\mathcal{F}_{RO}}(q_j)$ being the expected execution fidelity of a measurement operation~$q_j$ on its corresponding qubit and $|G|$ respectively $m$ being the number of gates and measurements in the compiled circuit.
\end{enumerate}
This figure of merit determines an estimate of the probability that a quantum circuit will return the expected result, the so-called \emph{expected fidelity}, which ranges between $0.0$ and $1.0$ with higher values being better.
\end{example}

\vspace{1cm}

However, the expected fidelity introduced above is just one exemplary figure of merit.
A potential alternative could be the \emph{critical depth} (taken from~\cite{supermarq})---a measure to describe the percentage of \mbox{multi-qubit} gates on the longest path through a compiled quantum circuit (determining the depth).
A respective value close to 1 would indicate a very sequential circuit while a value of 0 would indicate a highly parallel one.
On top of that, even combinations with customizable weighting are possible, such as $25\%$ of the expected fidelity and $75\%$ critical depth.
Furthermore, different figures of merit could be used for the device selection and the respective compilation itself.
While for the former, soft factors such as actual execution costs or queue lengths might be sensible to consider as well, the latter should only focus on technical factors that describe the quality of the compilation itself.

In the following, detailed descriptions of both the \mbox{RL-based} approach to the compilation and the \mbox{ML-based} approach to the device selection are given in \autoref{sec:rl} and \autoref{sec:ml}, respectively.
Afterwards, \autoref{sec:mqtpredictor} describes the combination of both techniques into a holistic framework.

\section{Compilation Using Reinforcement Learning}\label{sec:rl}
\begin{figure*}
\centering
\resizebox{0.99\linewidth}{!}{
\begin{tikzpicture}
\def\w2{1.0pt}
\def\minsize{1.3cm}
    \node (circ1)at ( -4.4, -3.5) [align=left]  {
\resizebox{0.22\linewidth}{!}{
	\begin{tikzpicture}
				  \begin{yquant}
				    	qubit {${q_2}$} q;
						qubit {${q_1}$} q[+1];
						qubit {${q_0}$} q[+1];

						[name=left]
				    	box {$X$} q[0];
				    	[name=right]
				    	box {$X$} q[0];

				    	box {$H$} q[1];
				    	box {$R_Z$} q[1];

				    	cnot q[1] | q[0];
				    	cnot q[2] | q[1];
				    	cnot q[2] | q[0];

				  \end{yquant}
				\end{tikzpicture} }
  };

  \node(points) at (-2.6,-3.5) {\huge{$...$}};
    \node[minimum width = 5.2cm, inner ysep=0mm, label=\textbf{Training Circuits}, draw, fit=(circ1)] (train){};

	\node (con_tikz)  [below of=train, yshift=-17mm, align=left] {
	Connectivity:\\
	   \begin{tikzpicture}
						    \node (p0) at ( 0, -1) [circle, draw]{$Q_0$};
						    \node (p1) at ( 1,-1) [circle, draw]{$Q_1$};
						    \node (p2) at ( 2,-1) [circle, draw]{$Q_2$};
						    \begin{scope}[every path/.style={-}]
						       \draw (p0) -- (p1);
						       \draw (p1) -- (p2);
						    \end{scope}
	\end{tikzpicture}};
   \node (gates) [below of=con_tikz, align=left]{Native Gates:\\ \{\emph{$R_Z$}, \emph{$\sqrt{X}$}, \emph{X}, \emph{CNOT}\}};
	\node (fidelities) [below of=gates, align=center, yshift=2mm]{\hspace{-10mm}Calibration:\\ \{1Q: $99.7\%$, 2Q: $98.2\%$, RO: $97.5\%$\}};

\node[minimum width = 5.2cm, inner ysep=0mm, label=\textbf{Quantum Device}, draw, fit=(con_tikz) (fidelities)] (dev){};

    \node (merit) [below of=dev, yshift=-25mm, align=left, xshift=0mm]  {
    Expected Fidelity: \\
    $\mathit{\mathcal{F}}=\prod_{i=1}^{|G|} \mathit{\mathcal{F}}(g_i) \prod_{j=1}^{m} \mathit{\mathcal{F}_{RO}}(q_j)$
  };

\node[right of=dev, yshift=22mm, xshift=10mm] {
\resizebox{0.06\linewidth}{!}{
\begin{tikzpicture}
        \node (p0) at ( 0, 0) [circle, fill]{};
        \node (p1) at ( 0, 1.0) [circle, fill]{};
        \node (p2) at ( -0.8, -0.8) [circle, fill]{};
        \node (p3) at ( -1, 0.5) [circle, fill]{};
        \node (p4) at ( 1, 0.5) [circle, fill]{};
        \node (p5) at ( 0.8, -0.8) [circle, fill]{};
           \draw[line width = 2] (p0) -- (p1);
           \draw[line width = 2] (p0) -- (p2);
           \draw[line width = 2] (p0) -- (p3);
           \draw[line width = 2] (p0) -- (p4);
           \draw[line width = 2] (p0) -- (p5);
    \end{tikzpicture}	}
    };
\node[minimum width = 5.2cm, inner ysep=0mm, label=\textbf{Figure of Merit}, draw, fit=(merit)] (fig_merit){};
\node[right of=train, yshift=12mm, xshift=11mm] {
\resizebox{0.1\linewidth}{!}{
\begin{tikzpicture}
     \node (qc) {
\begin{tikzpicture}
     \begin{yquant}
            qubit {\huge{${q_2}$}} q;
        qubit {\huge{${q_1}$}} q[+1];
        qubit {\huge{${q_0}$}} q[+1];
            box {} q[0];
            box {} (q[1], q[0]);
            box {} (q[1], q[2]);
      \end{yquant}
     \end{tikzpicture}
      };

     \node (points)[right of=qc, xshift=7mm] {\huge{$...$}};

    \end{tikzpicture}	}
    };
\node[right of=fig_merit, yshift=10mm, xshift=10mm] {
\resizebox{0.06\linewidth}{!}{
\begin{tikzpicture}
\draw [line width = 2] (0,0) arc (-360:-180:1cm);
\draw [-latex, line width = 2] (-1.0,0) -- (-0.2, 1.2);

    \end{tikzpicture}	}
    };
\def\height_mdp{-7.2}
    \def\w2{1.0pt}
\def\minsize{0.6cm}

	\node (p0) at ( 0, \height_mdp) [minimum size=\minsize,rectangle,rounded corners,draw,align=left]{\textcolor{red}{$\bm{\xmark}$} Nat. Gates\\\textcolor{red}{$\bm{\xmark}$} Topology};
    \node (p3) at ( 3,\height_mdp)  [minimum size=\minsize,rectangle,rounded corners,draw,align=left]{\textcolor{green}{$\bm{\cmark}$} Nat. Gates\\\textcolor{red}{$\bm{\xmark}$} Topology};
    \node (p4) at ( 6,\height_mdp) [minimum size=\minsize,rectangle,rounded corners, double,draw,align=left]{\textcolor{green}{$\bm{\cmark}$} Nat. Gates\\\textcolor{green}{$\bm{\cmark}$} Topology};

    \path[-latex] (p0) edge [] node [align=center, above, yshift=-2.5mm]  {\emph{Syn.} \\ /} (p3);
    \path[-latex] (p3) edge [] node [align=center, above, yshift=-2.5mm]  {\emph{Map.} \\ /} (p4);

    \def\lin{120}
    \def\lout{60}
    \def\lloose{7}
    \tikzset{every loop/.style={in=\lin,out=\lout,looseness=\lloose}}

    \path[-latex]  (p0) edge [dashed, loop above, line width=\w2] node [align=center, above, yshift=-2.5mm] {/}(p0);
    \path[-latex]  (p3) edge [dashed, loop above, line width=\w2] node [align=center, above, yshift=-2.5mm]{/}(p3);
    \path[-latex]  (p4) edge [dashed, loop above, line width=\w2] node [align=center, above, yshift=-2.5mm]{/}(p4);

\def\h{0.4}
\draw[-latex] (p4.270) [dashed, line width=\w2] .. controls ($(p4.south) - (0.1,\h)$) .. node [align=center, above, yshift=-3mm, xshift=-9mm]{/} ($(p4.south) - (1, \h)$) -- ($(p3.south) - (-1.5, \h)$) .. controls ($(p3.south) - (-1.2, \h)$) .. (p3.315);
\draw[-latex] (p3.270) [dashed, line width=\w2] .. controls ($(p3.south) - (0,\h)$) .. node [align=center, above, yshift=-3mm, xshift=-9mm]{/} ($(p3.south) - (1, \h)$) -- ($(p0.south) - (-1.5, \h)$) .. controls ($(p0.south) - (-1.2, \h)$) .. (p0.315);
\def\hd{0.55}
\draw[-latex] (p4.290) [dashed, line width=\w2] .. controls ($(p4.south) - (-0.1,\hd)$) .. node [align=center, above, yshift=-3mm, xshift=-28mm]{/} ($(p4.south) - (1, \hd)$) -- ($(p0.south) - (-1.2, \hd)$) .. controls ($(p0.south) - (0, \hd)$) .. (p0.270);
\node (desc) [text=black, below left=1.0cm and -1cm  of  p4] {multiple passes};
\path[-latex, line width=\w2] (desc.west) edge [] node [align=center, above, yshift=-2.4mm, xshift=1mm] {/} ($(desc.west) - (0.5, 0.0)$);
\node (desc2) [text=black, below left=1.3cm and -1cm  of  p4] {(multiple) opt. passes};
\path[-latex, line width=\w2] (desc2.west) edge [dashed] node [align=center, above, yshift=-2.4mm, xshift=1mm] {/} ($(desc2.west) - (0.5, 0.0)$);

\node[inner ysep=10mm, fit=(p0)(p4),draw, yshift=2mm] (rl_mdp){};

    \draw[-latex] (dev.east)  -| ([xshift=-5mm, yshift=0mm]rl_mdp.west) |- (rl_mdp.west);
    \draw[-latex] (train.east)  -| ([xshift=-5mm, yshift=0mm]rl_mdp.west) |- (rl_mdp.west);
    \draw[-latex] (fig_merit.east)  -| ([xshift=-5mm, yshift=0mm]rl_mdp.west) |- (rl_mdp.west);

\node[right of =rl_mdp,  xshift=6.0cm, draw, align=left, label=\textbf{Trained RL Model}] (rl_model){
\begin{tikzpicture}
  \node(a){
          \resizebox{0.1\linewidth}{!}{
\begin{tikzpicture}[baseline=(current bounding box.center)]
				  \begin{yquant}
				    	qubit {\huge{${q_2}$}} q;
						qubit {\huge{${q_1}$}} q[+1];
						qubit {\huge{${q_0}$}} q[+1];
				    	box {} q[0];
				    	box {} (q[0], q[1]);
				    	box {} (q[2], q[1]);
				  \end{yquant}

    \end{tikzpicture}	}

    };

\node(b)[below =of a, yshift=10mm] {

    \resizebox{0.07\linewidth}{!}{
	\begin{tikzpicture}
  \node (p0) at ( 0, 0) [circle, fill]{};
        \node (p1) at ( 0, 1.0) [circle, fill]{};
        \node (p2) at ( -0.8, -0.8) [circle, fill]{};
        \node (p3) at ( -1, 0.5) [circle, fill]{};
        \node (p4) at ( 1, 0.5) [circle, fill]{};
        \node (p5) at ( 0.8, -0.8) [circle, fill]{};
           \draw[line width = 2] (p0) -- (p1);
           \draw[line width = 2] (p0) -- (p2);
           \draw[line width = 2] (p0) -- (p3);
           \draw[line width = 2] (p0) -- (p4);
           \draw[line width = 2] (p0) -- (p5);
	\end{tikzpicture}
    }

    };

    \node(c)[below =of b, yshift=10mm] {
    \resizebox{0.06\linewidth}{!}{
    \begin{tikzpicture}
    \draw [line width = 2] (0,0) arc (-360:-180:1cm);
    \draw [-latex, line width = 2] (-1.0,0) -- (-0.2, 1.2);
    \end{tikzpicture}}

    };

    \node(d)[right=of b, yshift=0mm, xshift=-10mm] {
			\huge{$\rightarrow$}
    };

    \node(e)[right=of d, yshift=0mm, xshift=-14mm] {
    \resizebox{0.15\linewidth}{!}{
	\begin{tikzpicture}[xshift=30mm,baseline=(current bounding box.center),]
				  \begin{yquant}
				    	qubit {\huge{${Q_2}$}} q;
					qubit {\huge{${Q_1}$}} q[+1];
					qubit {\huge{${Q_0}$}} q[+1];
				    	box {} q[2];
				    	box {} (q[2], q[1]);
				    	box {} (q[0], q[1]);
				  \end{yquant}
				\end{tikzpicture} }};
	\end{tikzpicture}
	};
    \draw[-latex] (rl_mdp.east) -- (rl_model.west);

\end{tikzpicture}
}
	\subfloat[Input.\label{fig:rl_1} ]{\hspace{.3\linewidth}}
	\subfloat[MDP for quantum compilation.\label{fig:rl_2} ]{\hspace{.4\linewidth}}
	\subfloat[Output.\label{fig:rl_3} ]{\hspace{.3\linewidth}}
	
\vspace{-3mm}
\caption{Training process of an RL model for quantum compilation.}
\label{fig:mdp}
\end{figure*}
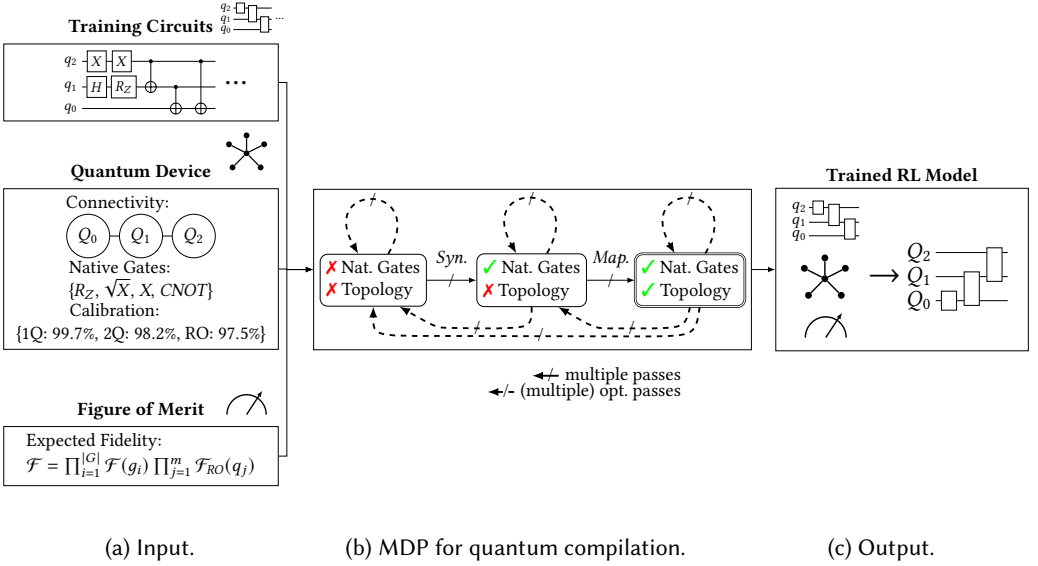

Compilation, fortunately, is not new \mbox{per-se}, since classical compilers have seen a similar trend of increasing complexity and variety in the past.
Instead of reinventing the wheel, we propose to
make use of the decades of research on classical compiler optimization
and model quantum circuit compilation in a similar fashion.
Based on that, classical reinforcement learning is used to learn optimized sequences of compilation passes for a given figure of merit.
The resulting model can then be used as a compiler.

\subsection{Quantum Circuit Compilation Modeled as a Markov Decision Process}\label{sec:mdp}
Quantum circuit compilation as described in \autoref{sec:compilation} can be interpreted as a \emph{Markov Decision Process} (MDP,~\cite{mdp}) that transforms a \mbox{high-level} circuit into an executable circuit by applying synthesis, mapping, and optimization passes.
For that, the compilation process is broken down into \emph{states} and corresponding \emph{actions} that can be executed in those states to advance the process.
Each state captures which of the requirements for a circuit to be executable are satisfied---specifically, whether the circuit only contains native gates and whether it matches the topology of the device.
This results in three states\footnote{We purposely omit the state where the circuit respects the topology but contains non-native gates as there are hardly any compilation passes available that take advantage of this specific combination of conditions.}:
\begin{enumerate}
	\item The circuit contains non-native gates and does not respect the device topology,
	\item The circuit only contains native gates but does not respect the device topology, and
	\item The circuit only contains native gates and respects the device topology, i.e., it is executable.
\end{enumerate}
The resulting MDP is depicted in \autoref{fig:rl_2}---with states being denoted as rounded rectangles and actions as arrows.
In the following, the individual states and actions will be described in more detail alongside an example.

Given a \mbox{high-level} quantum circuit and a target device (determining the native \mbox{gate-set} and qubit topology), compilation typically begins in the state where the circuit contains \mbox{non-native} gates and does not respect the device topology.
In this state, two kinds of actions are possible: Either a \emph{synthesis action} that transforms the circuit so that it only contains gates native to the target device, or an \emph{optimization action} that applies any kind of optimization to the circuit.

\begin{example}\label{ex:init}
To make this modeling more tangible, consider again the compilation flow shown in \autoref{fig:compflow} with the initial circuit depicted in \autoref{fig:sub_1}.
Since it comprises a \mbox{non-native} gate, the compilation starts from the \mbox{left-most} state in \autoref{fig:rl_2}.
First, an optimization pass is applied that cancels the two redundant \emph{X} gates.
Since the circuit still contains \mbox{non-native} gates, this does not change the overall state---as indicated by the dashed self-loop above the state.
Then, a synthesis pass is applied, which yields the circuit depicted in \autoref{fig:sub_2} and advances the state of the MDP to the next state.
\end{example}

Whenever the circuit only contains native gates but does not yet respect the device topology, three different actions are possible:
First, the circuit could be mapped using any \emph{mapping action}---making the circuit executable and, thus, advancing the MDP to the final state.
Additionally, the circuit may be optimized again.
Depending on whether the optimization pass preserves the native gates, the state after the action either stays the same (\emph{gate-set-preserving optimization action}) or goes back to the previous, non-native gates state (\emph{general optimization action}).

\begin{example}\label{ex:mdp_mapping}
Consider again the synthesized quantum circuit from \autoref{fig:sub_2}.
As before in \autoref{ex:init}, a similar optimization is conducted to cancel the two rotation gates.
Since the resulting circuit still only consists of native gates, this optimization does not change the state of the MDP---again represented by a dashed self-loop in \autoref{fig:rl_2}.
Subsequently, a mapping action is applied that leads to the quantum circuit depicted in \autoref{fig:sub_3} and advances the state of the MDP to the final state.
\end{example}

In the final state,
both the device's native \mbox{gate-set} and qubit connectivity constraints are fulfilled and any circuit in this state is already executable.
However, additional \emph{optimization} actions can be applied to further reduce complexity.
Again, those actions might violate any of the two constraints and the circuit might end up in a state where it is not executable any more.

\begin{example}
The quantum circuit resulting from \autoref{ex:mdp_mapping} is already executable.
However, another optimization action is applied to the circuit depicted in \autoref{fig:sub_3} to cancel two further gates and, by that, reducing the complexity.
Since this does not violate the device's connectivity, the circuit remains executable and the MDP state does not change---again represented by a dashed self-loop in \autoref{fig:rl_2}.
\end{example}

\vspace{2cm}

This MDP formulation enables a modular framework based on two properties:
\begin{enumerate}
	\item Each of the constraints that characterize an MDP state is \emph{efficiently computable} (via a single traversal of the circuit).
	\item All actions have a \emph{unified interface} based on a common intermediate representation (IR) for their input and output---independent of the origin of a certain action.
\end{enumerate}

The former ensures that, at any stage during the compilation, state transitions within the MDP can be computed efficiently.
The latter ensures interoperability between different SDKs and makes it possible
to \emph{mix and match} compilation passes provided by multiple SDKs.
Consequently, they can be combined and integrated in a single compilation framework---mitigating vendor \mbox{lock-ins}, allowing one to quickly adapt and integrate further compilation passes, and helping to keep up with the fast-paced development in quantum computing software.

\subsection{Learning Optimal Compilation Sequences Using Reinforcement Learning}
In classical compilation, different problems have been tackled very successfully using \emph{Reinforcement Learning} (RL)~\cite{reinforcement1,reinforcement2,reinforcement3}.
In general, these techniques aim to learn an \emph{action policy} that determines which actions should be applied based on \emph{observations} of the current state with the goal of maximizing a cumulative \emph{reward function}.
In contrast to providing labeled training data (as in \emph{Supervised Machine Learning}, ML), a \emph{training environment} representing the underlying MDP including its states and actions must be defined. 

To avoid reinventing the wheel, we also propose using such an RL approach for quantum circuit compilation based on the MDP shown in \autoref{fig:rl_2}.
To this end, three things must be provided as input (which are depicted in \autoref{fig:rl_1}):
\begin{enumerate}
\item \emph{Training Circuits}: Used in the training process to learn an action policy.
\item \emph{Quantum Device}: Defines the native \mbox{gate-set}, the qubit connectivity, and certain characteristics (such as gate fidelities) for the compilation process.
\item \emph{Figure of Merit}: Defines the evaluation score to guide the learning process.
\end{enumerate}
An \emph{RL agent} can then be employed to learn a corresponding action policy that maximizes the expected cumulative reward.
For the training itself, a \emph{sparse} reward function is used, i.e., as long as the circuit is not executable, the reward is always zero, and whenever the circuit becomes executable, the chosen figure of merit (e.g., expected fidelity as described in \autoref{sec:eval_metric}) is used to assign a score to the compiled circuit.

The resulting trained RL model then acts as \emph{compiler} itself that can compile any given quantum circuit for the target device it was trained for%
---as illustrated in \autoref{fig:rl_3}.
This creates a compiler that is fine-tuned for a specific device, optimizes for a customizable figure of merit, and factors in the input circuit when determining which compilation passes to apply---three major advances over current quantum circuit compilers.

\section{Device Selection Using Supervised Machine Learning}\label{sec:ml}

As discussed in \autoref{sec:compilation},
selecting the best quantum device for a particular application is itself a challenging task.
This section describes how that task can be interpreted as a classification problem and how supervised machine learning can predict the most promising qubit technology and device for a given quantum circuit---therefore, automatically make this decision for an \mbox{end-user}.

\subsection{Quantum Device Selection Modeled as a Classification Task}
A naive approach to selecting the best quantum device for a given quantum circuit would be to first compile it for all devices using a specific compiler, e.g., the trained RL models from \autoref{sec:rl}. %
Afterwards, the resulting compiled circuits must be evaluated according to some \emph{figure of merit} to identify the most promising device.
However, doing this for each and every \mbox{to-be-compiled} quantum circuit is practically infeasible since compilation is a \mbox{time-consuming} task and the number of available devices is steadily growing.

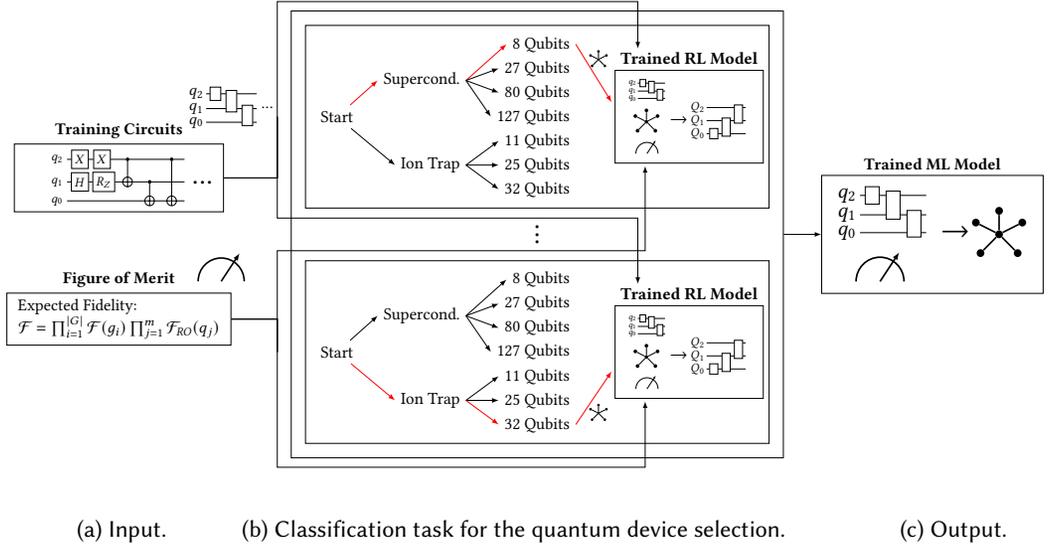
\begin{figure*}[t]
	\centering
	\resizebox{1.00\linewidth}{!}{
		\begin{tikzpicture}
			\node (qc) at ( 0, 0) [align=left]  {
				\resizebox{0.22\linewidth}{!}{
					\begin{tikzpicture}
						\begin{yquant}
							qubit {${q_2}$} q;
							qubit {${q_1}$} q[+1];
							qubit {${q_0}$} q[+1];
							box {$X$} q[0];
							box {$X$} q[0];
							box {$H$} q[1];
							box {$R_Z$} q[1];
							cnot q[1] | q[0];
							cnot q[2] | q[1];
							cnot q[2] | q[0];
						\end{yquant}
				\end{tikzpicture} }
			};
			\node(points) at (1.8,-0.1) {\huge{$...$}};

			\node (merit) [below of=qc, align=left, yshift=-20mm]  {
				Expected Fidelity: \\
				$\mathit{\mathcal{F}}=\prod_{i=1}^{|G|} \mathit{\mathcal{F}}(g_i) \prod_{j=1}^{m} \mathit{\mathcal{F}_{RO}}(q_j)$
			};

			\node[minimum width = 4.5cm, inner ysep=0mm, label=\textbf{Training Circuits}, draw, fit=(qc)] (train){};
			\node[minimum width = 4.5cm, inner ysep=0mm, label=\textbf{Figure of Merit}, draw, fit=(merit)] (fig_merit){};

			\node[right of=train, yshift=15mm, xshift=14mm] {
				\resizebox{0.15\linewidth}{!}{
					\begin{tikzpicture}
						\node (qc) {
							\begin{tikzpicture}
								\begin{yquant}
									qubit {\huge{${q_2}$}} q;
									qubit {\huge{${q_1}$}} q[+1];
									qubit {\huge{${q_0}$}} q[+1];
									box {} q[0];
									box {} (q[1], q[0]);
									box {} (q[1], q[2]);
								\end{yquant}
							\end{tikzpicture}
						};

						\node (points)[right of=qc, xshift=7mm] {\huge{$...$}};

				\end{tikzpicture}	}
			};

			\node[right of=fig_merit, yshift=11mm, xshift=12mm] {
				\resizebox{0.08\linewidth}{!}{
					\begin{tikzpicture}
						\draw [line width = 2] (0,0) arc (-360:-180:1cm);
						\draw [-latex, line width = 2] (-1.0,0) -- (-0.2, 1.2);

				\end{tikzpicture}	}
			};

			\node [right of = qc, xshift=8cm, yshift=1.3cm, draw] (top){

				\resizebox{0.7\linewidth}{!}{
					\begin{tikzpicture}
						\def\wdevs{0.0}
						\node (start) at (-5,-0.5)   {Start};
						\node[anchor=east] (superc) at (-2.3,0.25)   {Supercond.};
						\node[anchor=east] (ion) at (-2.3,-1.5)   {Ion Trap};
						\node[anchor=east] (8) at (\wdevs,1)   {8 Qubits};
						\node[anchor=east] (27) at (\wdevs,0.5)   {27 Qubits};
						\node[anchor=east] (80) at (\wdevs,0)   {80 Qubits};
						\node[anchor=east] (127) at (\wdevs,-0.5)   {127 Qubits};
						\node[anchor=east] (11) at (\wdevs,-1.0)   {11 Qubits};
						\node[anchor=east] (25) at (\wdevs,-1.5)   {25 Qubits};
						\node[anchor=east] (32) at (\wdevs,-2.0)   {32 Qubits};

						\node[anchor=east] (rl) at (1.0,-0.5)   {};

						\draw[-latex] (start) -- (ion.west);
						\draw[-latex, red] (start) -- (superc.west);
						\draw[-latex] (ion.east) -- (11.west);
						\draw[-latex] (ion.east) -- (25.west);
						\draw[-latex] (ion.east) -- (32.west);
						\draw[-latex, red] (superc.east) -- (8.west);
						\draw[-latex] (superc.east) -- (27.west);
						\draw[-latex] (superc.east) -- (80.west);
						\draw[-latex] (superc.east) -- (127.west);

						\draw[-latex, red] (8.east) -- node[above, black, -latex, every path/.style={-}, xshift=1mm] { \resizebox{0.03\linewidth}{!}{
								\begin{tikzpicture}
									\node (p0) at ( 0, 0) [circle, fill]{};
									\node (p1) at ( 0, 1.0) [circle, fill]{};
									\node (p2) at ( -0.8, -0.8) [circle, fill]{};
									\node (p3) at ( -1, 0.5) [circle, fill]{};
									\node (p4) at ( 1, 0.5) [circle, fill]{};
									\node (p5) at ( 0.8, -0.8) [circle, fill]{};
									\draw[line width = 2] (p0) -- (p1);
									\draw[line width = 2] (p0) -- (p2);
									\draw[line width = 2] (p0) -- (p3);
									\draw[line width = 2] (p0) -- (p4);
									\draw[line width = 2] (p0) -- (p5);
								\end{tikzpicture}
						}} ([xshift=0mm, yshift=3mm]rl.west);

						\node[right of =rl, draw, xshift=0.5cm, align=left, label=\textbf{Trained RL Model}] (rl_model){
							\resizebox{0.2\linewidth}{!}{

								\begin{tikzpicture}
									\node(a){
										\resizebox{0.1\linewidth}{!}{
											\begin{tikzpicture}[baseline=(current bounding box.center)]
												\begin{yquant}
													qubit {\huge{${q_2}$}} q;
													qubit {\huge{${q_1}$}} q[+1];
													qubit {\huge{${q_0}$}} q[+1];
													box {} q[0];
													box {} (q[0], q[1]);
													box {} (q[2], q[1]);
												\end{yquant}

										\end{tikzpicture}	}
									};

									\node(b)[below =of a, yshift=10mm] {
										\resizebox{0.07\linewidth}{!}{
											\begin{tikzpicture}
												\node (p0) at ( 0, 0) [circle, fill]{};
												\node (p1) at ( 0, 1.0) [circle, fill]{};
												\node (p2) at ( -0.8, -0.8) [circle, fill]{};
												\node (p3) at ( -1, 0.5) [circle, fill]{};
												\node (p4) at ( 1, 0.5) [circle, fill]{};
												\node (p5) at ( 0.8, -0.8) [circle, fill]{};
												\draw[line width = 2] (p0) -- (p1);
												\draw[line width = 2] (p0) -- (p2);
												\draw[line width = 2] (p0) -- (p3);
												\draw[line width = 2] (p0) -- (p4);
												\draw[line width = 2] (p0) -- (p5);
											\end{tikzpicture}
										}

									};

									\node(c)[below =of b, yshift=10mm] {
										\resizebox{0.06\linewidth}{!}{
											\begin{tikzpicture}
												\draw [line width = 2] (0,0) arc (-360:-180:1cm);
												\draw [-latex, line width = 2] (-1.0,0) -- (-0.2, 1.2);
										\end{tikzpicture}}
									};

									\node(d)[right=of b, yshift=0mm, xshift=-10mm] {
										\huge{$\rightarrow$}
									};

									\node(e)[right=of d, yshift=0mm, xshift=-14mm] {
										\resizebox{0.15\linewidth}{!}{
											\begin{tikzpicture}[xshift=30mm,baseline=(current bounding box.center),]
												\begin{yquant}
													qubit {\huge{${Q_2}$}} q;
													qubit {\huge{${Q_1}$}} q[+1];
													qubit {\huge{${Q_0}$}} q[+1];
													box {} q[2];
													box {} (q[2], q[1]);
													box {} (q[0], q[1]);
												\end{yquant}
									\end{tikzpicture} }};
								\end{tikzpicture}
							}

						};
				\end{tikzpicture}}
			};

			\node [below of = top, yshift=-4.0cm, draw] (bottom) {

				\resizebox{0.7\linewidth}{!}{
					\begin{tikzpicture}
						\def\wdevs{0.0}
						\node (start) at (-5,-0.5)   {Start};
						\node[anchor=east] (superc) at (-2.3,0.25)   {Supercond.};
						\node[anchor=east] (ion) at (-2.3,-1.5)   {Ion Trap};
						\node[anchor=east] (8) at (\wdevs,1)   {8 Qubits};
						\node[anchor=east] (27) at (\wdevs,0.5)   {27 Qubits};
						\node[anchor=east] (80) at (\wdevs,0)   {80 Qubits};
						\node[anchor=east] (127) at (\wdevs,-0.5)   {127 Qubits};
						\node[anchor=east] (11) at (\wdevs,-1.0)   {11 Qubits};
						\node[anchor=east] (25) at (\wdevs,-1.5)   {25 Qubits};
						\node[anchor=east] (32) at (\wdevs,-2.0)   {32 Qubits};

						\node[anchor=east] (rl) at (1.0,-0.5)   {};

						\draw[-latex, red] (start) -- (ion.west);
						\draw[-latex] (start) -- (superc.west);

						\draw[-latex] (ion.east) -- (11.west);
						\draw[-latex] (ion.east) -- (25.west);
						\draw[-latex, red] (ion.east) -- (32.west);
						\draw[-latex] (superc.east) -- (8.west);
						\draw[-latex] (superc.east) -- (27.west);
						\draw[-latex] (superc.east) -- (80.west);
						\draw[-latex] (superc.east) -- (127.west);

						\draw[-latex, red] (32.east) -- node[above, black, -latex, every path/.style={-}, xshift=1mm, yshift=-6mm] { \resizebox{0.03\linewidth}{!}{
								\begin{tikzpicture}
									\node (p0) at ( 0, 0) [circle, fill]{};
									\node (p1) at ( 0, 1.0) [circle, fill]{};
									\node (p2) at ( -0.8, -0.8) [circle, fill]{};
									\node (p3) at ( -1, 0.5) [circle, fill]{};
									\node (p4) at ( 1, 0.5) [circle, fill]{};
									\node (p5) at ( 0.8, -0.8) [circle, fill]{};
									\draw[line width = 2] (p0) -- (p1);
									\draw[line width = 2] (p0) -- (p2);
									\draw[line width = 2] (p0) -- (p3);
									\draw[line width = 2] (p0) -- (p4);
									\draw[line width = 2] (p0) -- (p5);
								\end{tikzpicture}
						}} ([xshift=0mm, yshift=-4mm]rl.west);

						\node[right of =rl, draw, xshift=0.5cm, align=left, label=\textbf{Trained RL Model}] (rl_model){
							\resizebox{0.2\linewidth}{!}{
								\begin{tikzpicture}
									\node(a){
										\resizebox{0.1\linewidth}{!}{
											\begin{tikzpicture}[baseline=(current bounding box.center)]
												\begin{yquant}
													qubit {\huge{${q_2}$}} q;
													qubit {\huge{${q_1}$}} q[+1];
													qubit {\huge{${q_0}$}} q[+1];
													box {} q[0];
													box {} (q[0], q[1]);
													box {} (q[2], q[1]);
												\end{yquant}

										\end{tikzpicture}	}
									};

									\node(b)[below =of a, yshift=10mm] {
										\resizebox{0.07\linewidth}{!}{
											\begin{tikzpicture}
												\node (p0) at ( 0, 0) [circle, fill]{};
												\node (p1) at ( 0, 1.0) [circle, fill]{};
												\node (p2) at ( -0.8, -0.8) [circle, fill]{};
												\node (p3) at ( -1, 0.5) [circle, fill]{};
												\node (p4) at ( 1, 0.5) [circle, fill]{};
												\node (p5) at ( 0.8, -0.8) [circle, fill]{};
												\draw[line width = 2] (p0) -- (p1);
												\draw[line width = 2] (p0) -- (p2);
												\draw[line width = 2] (p0) -- (p3);
												\draw[line width = 2] (p0) -- (p4);
												\draw[line width = 2] (p0) -- (p5);
											\end{tikzpicture}
										}

									};

									\node(c)[below =of b, yshift=10mm] {
										\resizebox{0.06\linewidth}{!}{
											\begin{tikzpicture}
												\draw [line width = 2] (0,0) arc (-360:-180:1cm);
												\draw [-latex, line width = 2] (-1.0,0) -- (-0.2, 1.2);
										\end{tikzpicture}}
									};

									\node(d)[right=of b, yshift=0mm, xshift=-10mm] {
										\huge{$\rightarrow$}
									};

									\node(e)[right=of d, yshift=0mm, xshift=-14mm] {
										\resizebox{0.15\linewidth}{!}{
											\begin{tikzpicture}[xshift=30mm,baseline=(current bounding box.center),]
												\begin{yquant}
													qubit {\huge{${Q_2}$}} q;
													qubit {\huge{${Q_1}$}} q[+1];
													qubit {\huge{${Q_0}$}} q[+1];
													box {} q[2];
													box {} (q[2], q[1]);
													box {} (q[0], q[1]);
												\end{yquant}
									\end{tikzpicture} }};
								\end{tikzpicture}

							}

						};
				\end{tikzpicture}}

			};

			\node[inner sep=3mm, fit=(bottom)(top),draw] (ml_class){};

			\draw[-latex] (train.east)  -| ([xshift=-6mm]top.west) |- ([xshift=0mm, yshift=5mm]top.42) --  ([xshift=0mm, yshift=-5mm]top.42);

			\draw[-latex] (train.east)  -| ([xshift=-6mm, yshift=0mm]top.west) |- ([xshift=0mm, yshift=9mm]bottom.42) --  ([xshift=0mm, yshift=-5mm]bottom.42);

			\node[align=center, rotate=90] at (9, -1.2) {\huge{$...$}};

			\draw[-latex] (fig_merit.east)  -| ([xshift=-6mm]bottom.west) |- ([xshift=0mm, yshift=-9mm]top.-40) --  ([xshift=0mm, yshift=9mm]top.-40);

			\draw[-latex] (fig_merit.east)  -| ([xshift=-6mm, yshift=0mm]bottom.west) |- ([xshift=0mm, yshift=-5mm]bottom.-40) --  ([xshift=0mm, yshift=9mm]bottom.-40);

			\node[right of =ml_class, draw, xshift=7.5cm, align=left, label=\textbf{Trained ML Model}] (ml_model){
				\begin{tikzpicture}
					\node(b) [yshift=3]{
						\resizebox{0.15\linewidth}{!}{
							\begin{tikzpicture}[baseline=(current bounding box.center)]
								\begin{yquant}
									qubit {\huge{${q_2}$}} q;
									qubit {\huge{${q_1}$}} q[+1];
									qubit {\huge{${q_0}$}} q[+1];
									box {} q[0];
									box {} (q[0], q[1]);
									box {} (q[2], q[1]);
								\end{yquant}

					\end{tikzpicture}	}};

					\node(c)[below =of b, yshift=10mm] {
						\resizebox{0.08\linewidth}{!}{
							\begin{tikzpicture}
								\draw [line width = 2] (0,0) arc (-360:-180:1cm);
								\draw [-latex, line width = 2] (-1.0,0) -- (-0.2, 1.2);
						\end{tikzpicture}}
					};

					\node(d)[right=of b, yshift=-4mm, xshift=-10mm] {
						\huge{$\rightarrow$}
					};

					\node(e)[right=of d, yshift=0mm, xshift=-14mm] {
						\resizebox{0.09\linewidth}{!}{
							\begin{tikzpicture}
								\node (p0) at ( 0, 0) [circle, fill]{};
								\node (p1) at ( 0, 1.0) [circle, fill]{};
								\node (p2) at ( -0.8, -0.8) [circle, fill]{};
								\node (p3) at ( -1, 0.5) [circle, fill]{};
								\node (p4) at ( 1, 0.5) [circle, fill]{};
								\node (p5) at ( 0.8, -0.8) [circle, fill]{};
								\draw[line width = 2] (p0) -- (p1);
								\draw[line width = 2] (p0) -- (p2);
								\draw[line width = 2] (p0) -- (p3);
								\draw[line width = 2] (p0) -- (p4);
								\draw[line width = 2] (p0) -- (p5);
							\end{tikzpicture}
					}};
				\end{tikzpicture}

			};
			\path[-latex] (ml_class.east) edge (ml_model.west);

		\end{tikzpicture}

	}
	\subfloat[Input.\label{fig:ml_1} ]{\hspace{.2\linewidth}}
	\subfloat[Classification task for the quantum device selection.\label{fig:ml_2} ]{\hspace{.55\linewidth}}
	\hspace{9mm}
	\subfloat[Output.\label{fig:ml_3} ]{\hspace{.15\linewidth}}
	\caption{Training process of an ML model for the quantum device selection.}
	\label{fig:training_data}
	\vspace{-4mm}
\end{figure*}

\begin{example}\label{ex:ml_1}
There are various underlying technologies how to realize quantum computers such as, e.g., superconducting- and ion \mbox{trap-based} qubits.
Even within a technology, respective devices may significantly vary in their characteristics such as the connectivity scheme, native \mbox{gate-set}, and fidelity rates.
Assuming that an \mbox{end-user} has access to seven devices based on two technologies, the task of determining the best device for a given circuit and figure of merit already becomes challenging since seven different compilations would have to be conducted---as insinuated in \autoref{fig:ml_2}.
\end{example}

Therefore, we propose an approach to \emph{learn} from previously conducted compilations and interpret the selection of the most promising device for a circuit and figure of merit as a statistical \emph{classification} task---a perfect fit for
supervised machine learning.
By that, \mbox{end-users} are freed from answering the following questions themselves:

\begin{itemize}
	\item Which \emph{qubit technology} is best suited for the application at hand?
	\item Which particular \emph{device}, e.g., from IBM or Rigetti, fits the quantum algorithm best?
\end{itemize}

\subsection{Predicting a Quantum Device Using Supervised Machine Learning}\label{sec:training_data}
\emph{Supervised Machine Learning} (ML) has proven a promising methodology in classical compilation (as described in~\cite{ml_in_compilers, autotuning_compiler}) but also in various other domains, such as, e.g., medicine, cyber security, and predictive analytics~\cite{sarkerMachineLearningAlgorithms2021}.
A crucial part to utilize this technology is gathering suitable training data which are representative for the whole problem domain space to facilitate generalizability---in this case, representative training circuits covering a broad range of applications.
Subsequently, each training circuit must be compiled for \emph{all} possible devices---using the trained RL models that act as compilers---to be able to identify the most promising according to the chosen figure of merit as shown in \autoref{fig:ml_2}.
That device then acts as the \emph{classification label} and, together with the training circuit itself, constitutes one \emph{training sample}.
To generate all the \emph{labeled training data} necessary to train respective ML models, this process is repeated for each training circuit and requires two things that must be provided (which are depicted in \autoref{fig:ml_1}):
\begin{enumerate}
\item \emph{Training Circuits}: Similar to the RL models, used here to generate labeled training data\footnote{To reduce the effort needed for regenerating labeled training data after changes, all compiled circuits including their evaluation score are persistently stored in a database.}.
\item \emph{Figure of Merit}: Determines the most promising device based on all conducted compilations. 
\end{enumerate}
Each labeled circuit must then be transformed into a \emph{feature vector} to be suitable for training a classifier.
There are many degrees of freedom when choosing the respective feature and it is very much an active field of research to determine features that best characterize a given quantum circuit.
In practice, these features should at least include the number of qubits and some statistics about the interactions of gates in the circuit, such as depth, parallelism, or interaction frequency.

Subsequently, the actual ML model can be trained.
This methodology with the respectively trained ML model (as illustrated in \autoref{fig:ml_3}) then acts as a tool to automatically predict the most promising quantum device for a given circuit and figure of merit. 
As a result, the user gets a recommendation that is circuit-specific, tuned for a specific figure of merit, and requires no compilation at all---again, three major advances over the established workflow.

\section{Proposed Approach: MQT~Predictor}\label{sec:mqtpredictor}
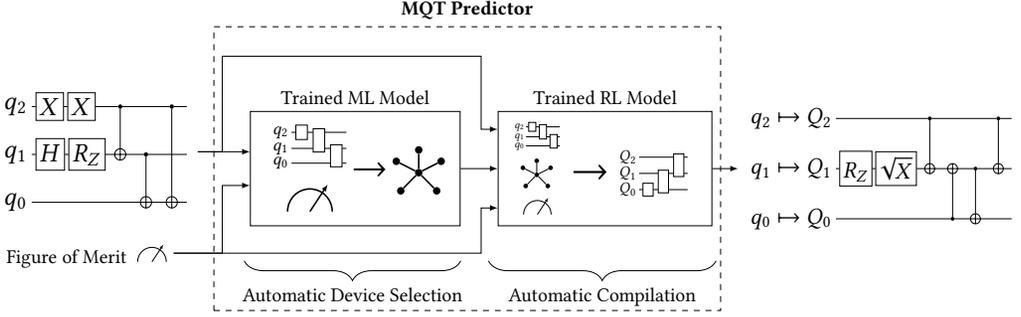
\begin{figure}
\centering
\resizebox{1.0\linewidth}{!}{
\begin{tikzpicture}
  \node (start) at (-6.0,-0.5)   {
\resizebox{0.25\linewidth}{!}{
\begin{tikzpicture}
				  \begin{yquant}[register/minimum height=1cm]
				    	qubit {\huge{${q_2}$}} q;
						qubit {\huge{${q_1}$}} q[+1];
						qubit {\huge{${q_0}$}} q[+1];

				    	box {\huge{$X$}} q[0];
				    	box {\huge{$X$}} q[0];

				    	box {\huge{$H$\vphantom{$R_Z$}}} q[1];
				    	box {\huge{$R_Z$}} q[1];

				    	cnot q[1] | q[0];
				    	cnot q[2] | q[1];
				    	cnot q[2] | q[0];

				  \end{yquant}
				\end{tikzpicture}  }
  };

\node[below=of start, label=Figure of Merit, xshift=-5mm] (merit_label){};
\node[right=of merit_label, yshift=5mm ](fig_merit){
    \resizebox{0.04\linewidth}{!}{
    \begin{tikzpicture}
    \draw [line width = 2] (0,0) arc (-360:-180:1cm);
    \draw [-latex, line width = 2] (-1.0,0) -- (-0.2, 1.2);
    \end{tikzpicture}}
};

\node[right of = start, draw, label=Trained ML Model, xshift=3.7cm, yshift=-3mm] (ml){
  \resizebox{!}{1.8cm}{
\begin{tikzpicture}

\node(b) [yshift=3]{
    \resizebox{0.1\linewidth}{!}{
\begin{tikzpicture}[baseline=(current bounding box.center)]
				  \begin{yquant}
				    	qubit {\huge{${q_2}$}} q;
						qubit {\huge{${q_1}$}} q[+1];
						qubit {\huge{${q_0}$}} q[+1];
				    	box {} q[0];
				    	box {} (q[0], q[1]);
				    	box {} (q[2], q[1]);
				  \end{yquant}

    \end{tikzpicture}	}};

    \node(c)[below =of b, yshift=10mm] {
    \resizebox{0.06\linewidth}{!}{
    \begin{tikzpicture}
    \draw [line width = 2] (0,0) arc (-360:-180:1cm);
    \draw [-latex, line width = 2] (-1.0,0) -- (-0.2, 1.2);
    \end{tikzpicture}}
    };

    \node(d)[right=of b, yshift=-4mm, xshift=-12mm] {
			\huge{$\rightarrow$}
    };

	  \node(e)[right=of d, yshift=0mm, xshift=-12mm] {
    \resizebox{0.07\linewidth}{!}{
	\begin{tikzpicture}
  \node (p0) at ( 0, 0) [circle, fill]{};
        \node (p1) at ( 0, 1.0) [circle, fill]{};
        \node (p2) at ( -0.8, -0.8) [circle, fill]{};
        \node (p3) at ( -1, 0.5) [circle, fill]{};
        \node (p4) at ( 1, 0.5) [circle, fill]{};
        \node (p5) at ( 0.8, -0.8) [circle, fill]{};
           \draw[line width = 2] (p0) -- (p1);
           \draw[line width = 2] (p0) -- (p2);
           \draw[line width = 2] (p0) -- (p3);
           \draw[line width = 2] (p0) -- (p4);
           \draw[line width = 2] (p0) -- (p5);
				\end{tikzpicture}
				}};
	\end{tikzpicture}}

	};

\node[right of = ml, align=left,draw,  label=Trained RL Model, xshift=3.5cm, minimum height=3] (rl){
  \resizebox{!}{1.8cm}{
\begin{tikzpicture}
  \node(a){
    \resizebox{0.1\linewidth}{!}{
\begin{tikzpicture}[baseline=(current bounding box.center)]
				  \begin{yquant}
				    	qubit {\huge{${q_2}$}} q;
						qubit {\huge{${q_1}$}} q[+1];
						qubit {\huge{${q_0}$}} q[+1];
				    	box {} q[0];
				    	box {} (q[0], q[1]);
				    	box {} (q[2], q[1]);
				  \end{yquant}

    \end{tikzpicture}	}
    };

\node(b)[below =of a, yshift=10mm] {
    \resizebox{0.07\linewidth}{!}{
	\begin{tikzpicture}
  \node (p0) at ( 0, 0) [circle, fill]{};
        \node (p1) at ( 0, 1.0) [circle, fill]{};
        \node (p2) at ( -0.8, -0.8) [circle, fill]{};
        \node (p3) at ( -1, 0.5) [circle, fill]{};
        \node (p4) at ( 1, 0.5) [circle, fill]{};
        \node (p5) at ( 0.8, -0.8) [circle, fill]{};
           \draw[line width = 2] (p0) -- (p1);
           \draw[line width = 2] (p0) -- (p2);
           \draw[line width = 2] (p0) -- (p3);
           \draw[line width = 2] (p0) -- (p4);
           \draw[line width = 2] (p0) -- (p5);
	\end{tikzpicture}
    }

    };

    \node(c)[below =of b, yshift=10mm] {
    \resizebox{0.06\linewidth}{!}{
    \begin{tikzpicture}
    \draw [line width = 2] (0,0) arc (-360:-180:1cm);
    \draw [-latex, line width = 2] (-1.0,0) -- (-0.2, 1.2);
    \end{tikzpicture}}
    };

    \node(d)[right=of b, yshift=0mm, xshift=-10mm] {

    \resizebox{0.105\linewidth}{!}{
			\huge{$\rightarrow$}
			}
    };

    \node(e)[right=of d, yshift=0mm, xshift=-14mm] {
    \resizebox{0.15\linewidth}{!}{
	\begin{tikzpicture}[xshift=30mm,baseline=(current bounding box.center),]
				  \begin{yquant}
				    	qubit {\huge{${Q_2}$}} q;
					qubit {\huge{${Q_1}$}} q[+1];
					qubit {\huge{${Q_0}$}} q[+1];
				    	box {} q[2];
				    	box {} (q[2], q[1]);
				    	box {} (q[0], q[1]);
				  \end{yquant}
				\end{tikzpicture} }
				};
	\end{tikzpicture}

				}
	};
\node[right of = rl, align=left,  xshift=4cm] (done){

\resizebox{0.35\linewidth}{!}{
	\begin{tikzpicture}
				  \begin{yquant}[register/minimum height=1.2cm]
            qubit {\huge{${q_2} \mapsto Q_2$}} q;
		qubit {\huge{${q_1} \mapsto Q_1$}} q[+1];
		qubit {\huge{${q_0} \mapsto Q_0$}} q[+1];
		box {\huge{$R_Z$\vphantom{$\sqrt{X}$}}} q[1];
		box {\huge{$\sqrt{X}$}} q[1];

	    cnot q[1] | q[0];
		cnot q[1] | q[2];
		cnot q[2] | q[1];
		cnot q[1] | q[0];

				  \end{yquant}
				\end{tikzpicture}
}
};

\def\warcs{-2.3}
\draw [decorate,decoration={brace,amplitude=10pt,mirror,raise=4pt},yshift=0mm]
(-3.3,\warcs) -- (0.6,\warcs) node [black,midway,below, yshift=-5mm] {Automatic Device Selection};
\draw [decorate,decoration={brace,amplitude=10pt,mirror,raise=4pt},yshift=0mm]
(1.1,\warcs) -- (5.2,\warcs) node [black,midway,below, yshift=-5mm] {Automatic Compilation};

  \draw[-latex] (rl.east) -- (done.west);
  \draw[-latex] (ml.east) -- (rl.west);

        \draw[-latex] (start.east)  -| ([xshift=-5mm, yshift=3mm]ml.west) |- ([xshift=0mm, yshift=3mm]ml.west);

  	\draw[-latex] (start.east)  -| ([xshift=-5mm, yshift=20mm]ml.west) |- ([xshift=-3mm, yshift=20mm]rl.west) -- ([xshift=-3mm, yshift=7mm]rl.west) -- ([xshift=0mm, yshift=7mm]rl.west);

      \draw[-latex] (fig_merit.east)  -| ([xshift=-5mm, yshift=-3mm]ml.west) |- ([xshift=0mm, yshift=-3mm]ml.west);

	\draw[-latex] (fig_merit.east)  -| ([xshift=-3mm, yshift=-7mm]rl.west) |- ([xshift=0mm, yshift=-7mm]rl.west);

\node[fit=(ml)(rl), draw, dashed, inner ysep=15mm, inner xsep=4mm, label = \textbf{MQT~Predictor}, xshift=-2.5mm] {};

\end{tikzpicture}

  }
	\caption{Automatic compilation flow selection and execution using MQT~Predictor.}
	\label{fig:combined_approach}
	\vspace{-4mm}
\end{figure}

Together with the trained RL models from \autoref{sec:rl}, the trained ML model automatically solves the device selection and compilation process---forming a holistic framework.
In the following, this framework, called the \emph{MQT~Predictor}, and its implementation are described in more detail.

\subsection{Approach}
The proposed approach is shown in \autoref{fig:combined_approach}.
It takes a quantum circuit and a customizable figure of merit as input.
Then, the framework itself predicts which quantum device is the most promising using the trained ML model---in \emph{realtime} without conducting any compilation at all.
The outcome of this is the predicted device.
Based on that, the respectively trained RL model compiles the given circuit for the predicted device optimizing for the chosen figure of merit.
Finally, it returns the compiled quantum circuit.
By that, the MQT~Predictor completely frees \mbox{end-users} from the daunting tasks of selecting a proper quantum device and finding a good compiler and can be used with just a single \emph{Python} function call---providing the same ease of use as the \mbox{state-of-the-art} compilers such as Qiskit~\cite{qiskit}.
On top of that, it combines the advantages of both individual techniques into a single powerful tool.

\begin{example}
Consider again the original scenario shown in \autoref{fig:search_space}, where an \mbox{end-user} wants to execute the circuit shown again on the \mbox{left-hand} side of \autoref{fig:combined_approach}.
Instead of manually finding the most promising device and compiler passes, the circuit can be simply fed into the MQT~Predictor framework together with the expected fidelity (as defined in \autoref{sec:eval_metric}) as its figure of merit.
Then, the MQT~Predictor predicts a respective device and compiles the circuit accordingly using the trained RL model---returning the compiled circuit shown on the \mbox{right-hand} side of \autoref{fig:combined_approach}.
\end{example}

The following sections describe the particular instantiation of the framework that has been implemented as part of this work.

\subsection{Implementation of Automatic Compilation Using RL}
The training environment described in \autoref{sec:rl} and shown in \autoref{fig:rl_2} is built on top of the \mbox{open-source} library \emph{OpenAI Gym}~\cite{openaigym}.
For that, all compilation passes from different sources that should be considered in learning promising compilation sequences must be implemented through a \emph{unified interface}.

To this end, \emph{Qiskit's QuantumCircuit} is used as the \emph{intermediate representation}.
Therefore, a compilation pass can be incorporated if the following requirement is fulfilled:
Either it must provide an interface for both the input and the compiled output circuit that supports \emph{Qiskit's QuantumCircuit} or a parser that supports \emph{OpenQASM}~(version $2$ or $3$)---which, then, can be used to create and serialize a \emph{Qiskit's QuantumCircuit} again.

As modeled in \autoref{fig:mdp}, three different types of actions are considered: synthesis, mapping, and optimization passes.
Using the described unified interface for both input and output of a compilation action, compilation passes from various quantum SDKs can easily be incorporated.
To demonstrate this, representative compilation passes from IBM's Qiskit (version $0.43.3$) and Quantinuum's TKET (version $1.17.1$) have been integrated---leading to the following list of available actions.

\begin{itemize}
	\item \emph{Synthesis}: Qiskit's \emph{BasisTranslator}
	\item \emph{Mapping}: Qiskit's \emph{Sabre} mapping or any combination of the following methods
	\begin{itemize}
		\item Layout:
		\begin{itemize}
			\item Qiskit's \emph{TrivialLayout}
			\item Qiskit's \emph{DenseLayout}
			\item Qiskit's \emph{SabreLayout}
		\end{itemize}
		\item Routing:
		\begin{itemize}
			\item Qiskit's \emph{BasicSwap}
			\item Qiskit's \emph{StochasticSwap}
			\item Qiskit's \emph{SabreSwap}
			\item TKET's \emph{RoutingPass}
		\end{itemize}
	\end{itemize}
	\item \emph{Optimization}:
	\begin{itemize}
		\item Qiskit's \emph{Optimize1qGatesDecomposition}
		\item Qiskit's \emph{CXCancellation}
		\item Qiskit's \emph{CommutativeCancellation}
		\item Qiskit's \emph{CommutativeInverseCancellation}
		\item Qiskit's \emph{RemoveDiagonalGatesBeforeMeasure}
		\item Qiskit's \emph{InverseCancellation}
		\item Qiskit's \emph{OptimizeCliffords}
		\item Qiskit's \emph{Collect2qBlocks} + \emph{ConsolidateBlocks}
		\item Qiskit's \emph{O3 Fixpoint Optimization Routine}
		\item TKET's \emph{PeepholeOptimise2Q}
		\item TKET's \emph{CliffordSimp}
		\item TKET's \emph{FullPeepholeOptimise}
		\item TKET's \emph{RemoveRedundancies}
	\end{itemize}
\end{itemize}

For the training process, more than $500$ training circuits from \emph{MQT~Bench}~\cite{quetschlich2023mqtbench} on the target-independent level have been used ranging from $2$ to $30$ qubits.
Lastly, a maskable version of the \emph{Proximal Policy Optimization (PPO)} algorithm~\cite{ppo_algo} provided by \emph{\mbox{Stable-Baselines3}}~\cite{stable-baselines3} is used for the reinforcement learning training process itself and also takes care of potential infinite sequences due to the backward loops within the MDP.

\subsection{Implementation of Automatic Device Selection Using ML}
For the quantum device selection, the ML model described in \autoref{sec:ml} and shown in \autoref{fig:ml_2} is trained using \mbox{scikit-learn}~\cite{scikit-learn}.
In this regard, seven different underlying algorithms are implemented and evaluated in~\cite{quetschlich2023prediction}---with \mbox{grid-searched} and \mbox{5-fold} \mbox{cross-validated} parameter values---to determine the most promising one:
\begin{itemize}
\item \emph{Random Forest} \cite{breiman2001random}
\item \emph{Gradient Boosting} \cite{friedman2001greedy}
\item \emph{Decision Tree} \cite{BreiFrieStonOlsh84}
\item \emph{Nearest Neighbor} \cite{cover1967nearest}
\item \emph{Multilayer Perceptron} \cite{haykin1994neural}
\item \emph{Support Vector Machine} \cite{cortes1995support}
\item \emph{Naive Bayes} \cite{bayes}
\end{itemize}
In~\cite{kotsiantis2007supervised}, an overview of today's use of those techniques is given.
All seven machine learning classifiers have been experimentally evaluated.
Since the \emph{Random Forest} classifier has given the best results, it is used in the following as the ML algorithm.

As indicated in \autoref{fig:search_space}, seven devices from two qubit technologies are considered as representatives:
\begin{itemize}
	\item Superconducting: four devices with $8$, $27$, $80$, and $127$ qubits
	\item Ion Trap: three devices with $11$, $25$, and $32$ qubits
\end{itemize}

To generate respective training data, more than $500$ training circuits from \emph{MQT~Bench}~\cite{quetschlich2023mqtbench} on the \mbox{target-independent} level are used (ranging from $2$ to $90$ qubits) with a $70/30$ \mbox{train-test-split}.
The compilations to generate the labeled training data are conducted based on the trained RL models acting as respective compilers.

\subsection{Quantum Circuit Representation}
Both the RL and the ML training process require quantum circuits to be represented as \mbox{so-called} \emph{feature vectors}--a vectorized representation based on integer and floating point values that makes them suitable for training a respective model.
Similar to the figure of merit described in \autoref{sec:eval_metric}, these features can be arbitrarily complex.
In this instantiation, various characteristics are used to describe a quantum circuit for both models: the \emph{number of qubits}, the \emph{depth} of the circuit, and the five composite features of \emph{program communication}, \emph{\mbox{critical-depth}}, \emph{\mbox{entanglement-ratio}}, \emph{parallelism}, and \emph{liveness} originally proposed in~\cite{supermarq}:

\begin{itemize}
	\item \emph{Program Communication}: Metric to measure the average degree of interaction for all qubits. A value of $1$ indicates that each qubit interacts at least once with all other qubits.
	\item \emph{Critical Depth}: Metric to measure how many of all \mbox{multi-qubit} gates are on the longest path (defining the depth of a quantum circuit).
	A value of $1$ indicates that all \mbox{multi-qubit} gates are on the longest path.
	\item \emph{Entanglement Ratio}: Metric to measure how many gates in a quantum circuit are \mbox{multi-qubit} gates. A value of $1$ indicates that the quantum circuit consists of only \mbox{multi-qubit} gates.
	\item \emph{Parallelism}: Metric to measure how much parallelization within the circuit is possible due to simultaneous gate execution. A value of $1$ describes large parallelization.
	\item \emph{Liveness}: Metric to measure how often the qubits are idling and waiting for their next gate execution. A value of $1$ describes a circuit in which there is a gate execution on each qubit at each time step.
\end{itemize}

Furthermore, for the ML model, the number of gates for each gate type according to the \emph{OpenQASM 2.0} specification~\cite{crossOpenQuantumAssembly2017} are used as features.
For the RL model, this was omitted to improve the learning efficiency, since the gate count features have been shown to be less representative than the composite ones~\cite{quetschlich2023prediction}.

\section{Evaluation}\label{sec:evaluation}
Based on the instantiation described above, this section demonstrates the feasibility of the MQT~Predictor framework by means of numerical evaluation.
All used \mbox{pre-trained} models and classifiers are publicly available as \mbox{open-source} on GitHub (\url{https://github.com/cda-tum/mqt-predictor}) and as an \mbox{easy-to-use} \emph{Python} package (\url{https://pypi.org/p/mqt.predictor}).

\subsection{Setup}
To assess the quality of the proposed framework, it is evaluated on more than $150$ benchmarks from $2$ to $90$ qubits taken from \emph{MQT~Bench}~\cite{quetschlich2023mqtbench} using the \mbox{target-independent} level circuits and the MQT~Predictor instantiation described in \autoref{sec:mqtpredictor}.
For each of those evaluation circuits---which have not been part of the (ML) training circuits---the tool predicts the most promising target device and its optimized compiler to compile the given circuit to the selected device.
To assess quality, two figures of merit are used:
\begin{enumerate}
\item The expected fidelity as introduced in~\autoref{sec:eval_metric} and
\item The critical depth describing the ratio of \mbox{multi-qubit} gates on the longest path of the compiled quantum circuit\footnote{Those ratios are actually subtracted from $1$ such that higher values indicate a more parallel quantum circuit that is assumed to be better.}.
\end{enumerate}
Note that the used figures of merit are examples and the MQT~Predictor framework allows \mbox{end-user} to provide \emph{any} metric to optimize for. 
In this case, the focus rather lies on technical factors to determine both the best device and compilation sequence while it might be sensible to also factor in soft factors such as the execution costs or the queue length for the device selection.
Since the focus of this work is to provide a framework that optimizes for \emph{any} given figure of merit, the following evaluations focus on how well the framework optimizes for these exemplary figures of merit and not on how suited these figures of merit are to assess the compilation task itself.

To generate baseline values, all evaluation circuits are compiled for all seven currently supported devices using both Qiskit's and TKET's most optimized (\mbox{pre-configured and fixed}) compilation settings \emph{O3} respectively \emph{O2}---leading to $14$ different compiled circuits as baselines.

\subsection{Results}\label{sec:results}
\begin{figure*}[t]
	\begin{center}
		\includegraphics[width=0.99\linewidth]{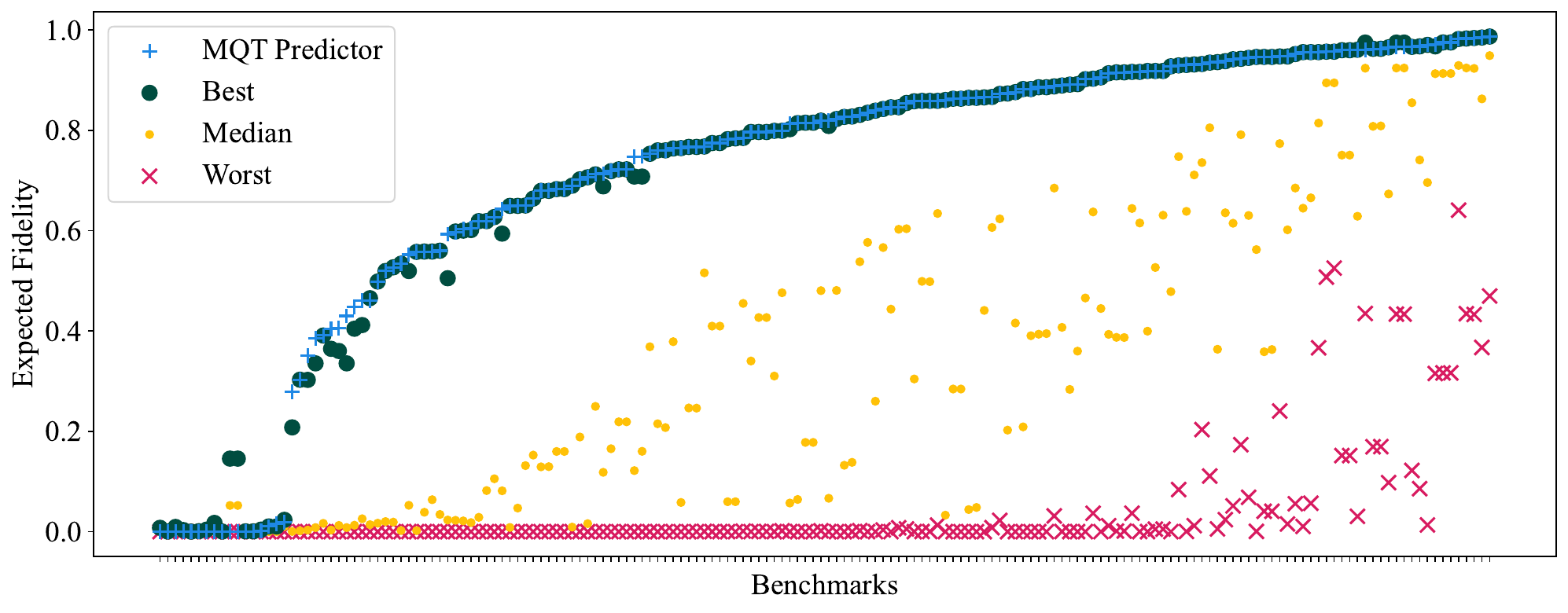}
	\end{center}
	\caption{Evaluation of the expected fidelity. All benchmarks are sorted by their MQT~Predictor score for better readability.}
	\label{fig:fidelity_plot}
	\vspace{-3mm}
\end{figure*}

\autoref{fig:fidelity_plot} shows the results obtained by using the proposed framework to optimize the expected fidelity.
Each tick on the \mbox{x-axis} denotes one evaluation circuit that is assessed using the $14$ baseline compilation flows ($7$ devices $\times~2$ compilers) and the MQT~Predictor.
Benchmarks leading to only zero values have been excluded from all evaluations.
Since (maximally) $15$ values per tick would be hard to read, the graph is reduced to only show the
\begin{itemize}
\item MQT~Predictor evaluation score in blue,
\item best evaluation score in green,
\item median evaluation score in yellow, and
\item worst evaluation score in red.
\end{itemize}

In this setup, the proposed method is capable of producing compiled quantum circuits that are within the \mbox{top-3} of all $14$ baselines in over $98\%$ of cases while frequently outperforming all baselines by up to $53\%$.
Therefore, especially \mbox{end-users} who are not experts in quantum computing can use it to reliably and automatically compile quantum circuits whose quality otherwise can only be achieved by many manual experiments---often infeasible due to the necessary time effort and expert knowledge.
These results also clearly highlight that choosing the right device and compiler can make the difference between a successful execution and obtaining completely random results.

\begin{figure*}[t]
	\begin{center}
		\includegraphics[width=0.99\linewidth]{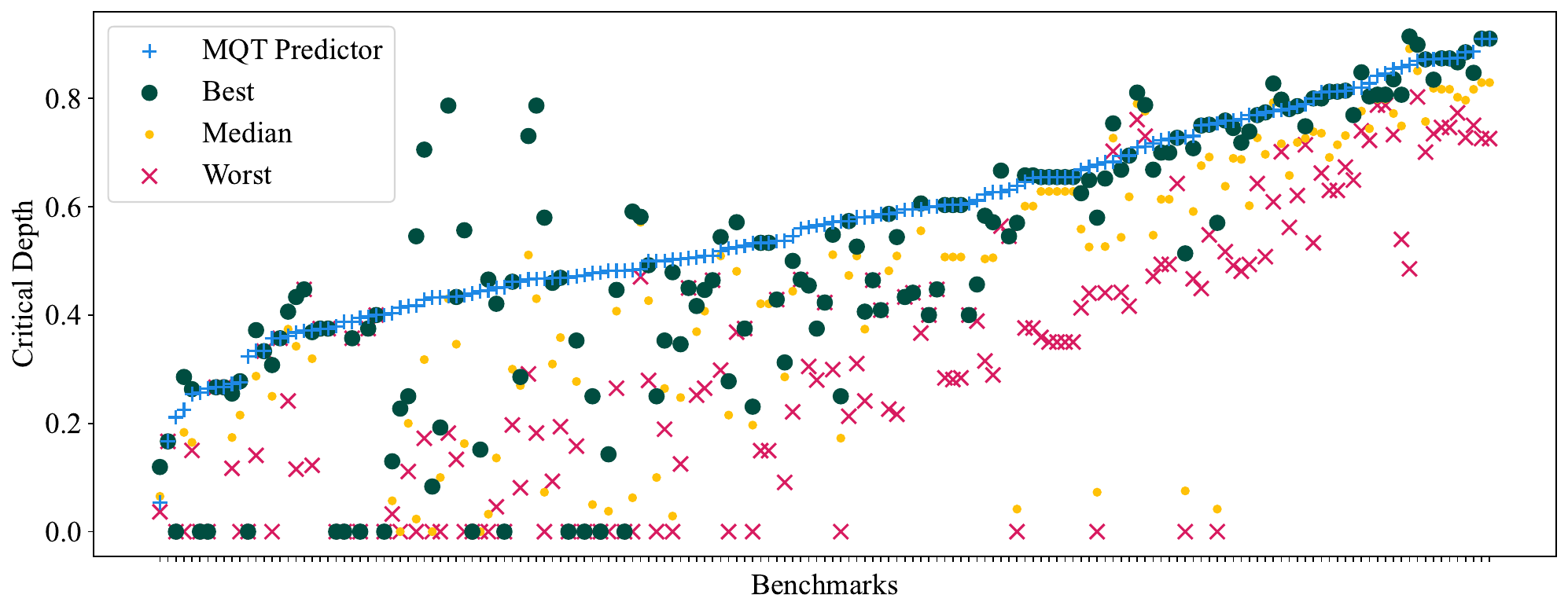}
	\end{center}
	\caption{Evaluation of the critical depth. All benchmarks are sorted by their MQT~Predictor score for better readability.}
	\label{fig:depth_plot}
	\vspace{-3mm}
\end{figure*}

To demonstrate the ability to optimize for customizable figures of merit, we additionally set up the framework to optimize for critical depth and show the respective results in \autoref{fig:depth_plot}.
Again, benchmarks only leading to zero values have been excluded from the evaluations.
Since it is not possible to define a custom figure of merit/optimization criterion for both Qiskit and TKET, the same compiled baseline evaluation circuits are used.
The respective graph has been reduced in a similar fashion as described above.
For $89\%$ of cases, the compiled circuits produced by MQT~Predictor are within the \mbox{top-3} of the $14$ baselines while frequently outperforming all baselines by up to $400\%$.

\begin{table}[t]
\caption{Comparison of the influence of the figure of merit.}
\label{tab:comparison}
\centering
\resizebox{.5\linewidth}{!}{
\begin{tabular}{l|cc}
& \multicolumn{2}{c}{Average result for...} \\
Model trained for... & Exp. Fidelity  & Critical depth \\
\hline
Exp. Fidelity & \textbf{0.67}  & 0.42 \\
Critical depth & 0.12   & \textbf{0.55} \\
\end{tabular}
}
\vspace{-5mm}
\end{table}

These results describe the performance from the \mbox{end-users'} perspective, who use the MQT~Predictor framework to automatically select the most promising device and compile accordingly. 
To provide more insight into the underlying training data, the ideal device distribution for the evaluated benchmarks is explored in \aref{appendix:dev_dist}.
Additionally, the impact of both the device selection as well as the compilation has been examined in an isolated fashion in \aref{appendix:ml} and \aref{appendix:rl}, respectively.

To further investigate the frameworks ability to optimize for customizable figures of merit, the average evaluation scores for both figures of merit are calculated for both respectively trained frameworks. 
The resulting numbers are denoted in \autoref{tab:comparison} and confirm that the framework trained for a particular figure of merit indeed results in the compiled circuits with the highest evaluation scores for both cases.

\subsection{Generalizability}
The performance of machine learning models, in general, is highly dependent on the quality of the training data.
Therefore, training data should be as heterogeneous as possible to cover a wide variety of circuits with different characteristics to achieve a high \emph{generalizability}.
For that purpose, a multitude of different algorithms are considered in the training data for both the ML and RL models described above---ranging from quantum circuit building blocks such as, e.g., the \emph{quantum fourier transform} and \emph{amplitude estimation} towards rather complex quantum circuits such as, e.g., ground state estimation---all taken from \emph{MQT~Bench}~\cite{quetschlich2023mqtbench}.

\begin{figure*}[t]
     \begin{subfigure}[b]{0.40\textwidth}
         \centering
         \includegraphics[width=\textwidth]{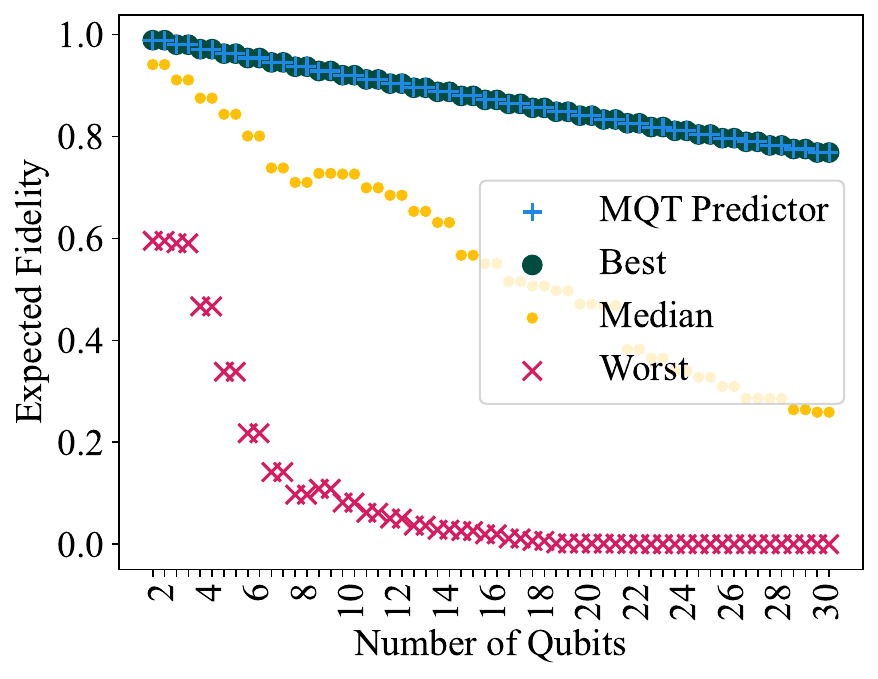}
         \caption{Expected fidelity for GHZ circuits.}
         \label{fig:fid_ghz}
     \end{subfigure}
     \hfill
     \begin{subfigure}[b]{0.40\textwidth}
         \centering
         \includegraphics[width=\textwidth]{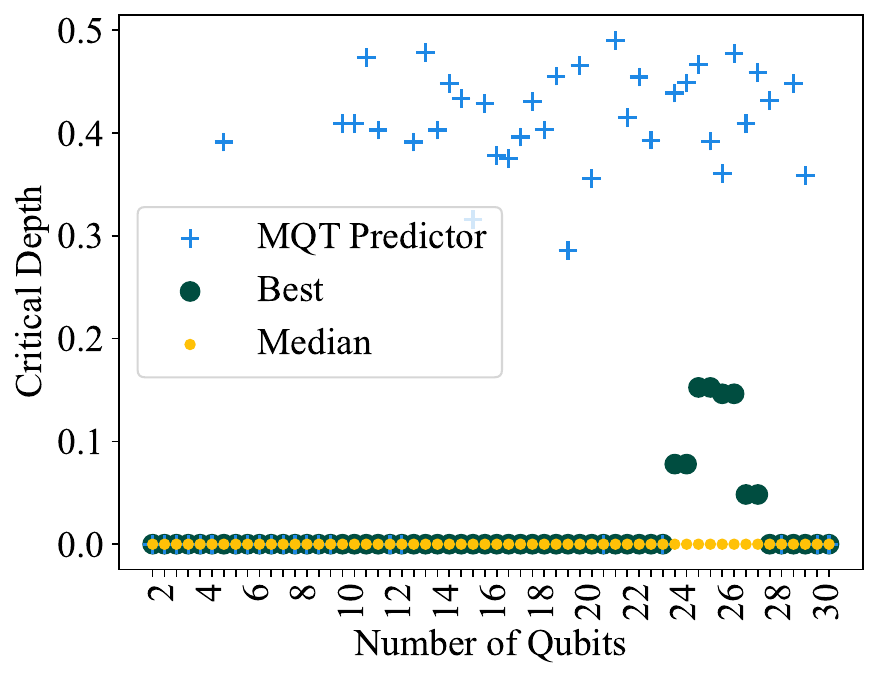}
         \caption{Critical depth for GHZ circuits.}
         \label{fig:depth_ghz}
     \end{subfigure}
\caption{Generalizability evaluation for GHZ circuits which were not part of the training processes.}
\vspace{-5mm}
\end{figure*}

To estimate how well the trained MQT~Predictor works for not only the quantum circuits that are part of the training data but also for other algorithms that have not been considered during training, a further evaluation has been conducted.
To evaluate generalizability, quantum circuits for the \emph{Greenberger–Horne–Zeilinger} (GHZ) state, which have not been considered during training, are evaluated in a similar fashion as before in \autoref{sec:results}.
The resulting scores for expected fidelity and critical depth are denoted in \autoref{fig:fid_ghz} and \autoref{fig:depth_ghz}, respectively.

Taking into account both figures of merit, a performance similar to the results described in \autoref{sec:results} is achieved.
Considering the expected fidelity, MQT~Predictor leads to the same performance as the best baseline for all cases.
In case of the critical depth as a figure of merit, MQT~Predictor even outperforms all baselines in most cases.
It should be noted that the best baseline results for critical depth frequently are $0.0$ (since \emph{all} \mbox{two-qubit} gates are on the longest path), while the MQT~Predictor achieves significant improvements in many of these cases.
Although this is only a small evaluation of the generalizability of the proposed methodology, it indicates that it also works well in untrained scenarios.

\section{Discussion}
\label{sec:discussion}
Classical machine learning has shown tremendous capabilities to solve problems from various domains such as classical compilation and beyond.
However, its potential success inherently depends on gathering a sufficiently large amount of high-quality training data and, by that, it is not \mbox{per-se} well suited for changes, which often require to \mbox{re-train} the underlying machine learning models.
This section specifically gives guidance how the MQT~Predictor deals with such changes in various ways.

\subsection{Changes in Hardware Information used in Figure of Merit}
Quantum computers are frequently calibrated to maximize the actual gate fidelities, e.g., used in the figure of merit introduced in \autoref{sec:eval_metric}.
Thus, these calibration updates affect the fidelity data used throughout both the RL and ML training as part of the cost function.
Consequently, the models need to be \mbox{re-trained}.
Unfortunately, it is not feasible to always generate all training data from scratch whenever such a calibration is conducted since it may occur on a daily (or even hourly) basis, while the training of the proposed framework instantiation took approximately $24$ hours on a consumer MacBook Pro.
As with any AI solution currently, a natural way to speed up the training time would be to utilize supercomputers, e.g., using \emph{High Performance Computing} clusters.
Future work may focus on machine learning techniques to adjust existing models to new circumstances that do not involve \mbox{re-training} from scratch.

\subsection{Additional Compilation Passes and Figures of Merit}
The quantum software ecosystem is moving fast and quantum SDKs frequently release new versions of their toolchains.
In addition, many new research methods for quantum circuit compilation are proposed regularly.
In order to keep up with this pace, the MQT~Predictor has been designed with a unified and modular interface that allows one to easily add and/or update compilation passes.
The good news is that the degree of change---whether just another optimization pass has been added or whether all of it changed---does not affect the overall effort.
The bad news is that any such change still requires \mbox{re-training} of all RL and ML models.
Also here, future work may focus on finding ways to \mbox{re-train} models without starting from scratch.

A similar effort is necessary when changing or adapting the figure of merit.
Since both the RL and the ML models optimize for it during training, again, the whole MQT~Predictor framework must be \mbox{re-trained} and the persistently stored data must be refreshed.

\subsection{Changes in Supported Quantum Devices}
New quantum devices become available on a regular basis.
Fortunately, adding a new device to the proposed framework is rather easy.
It is only necessary to train the respective RL model and to determine the respective evaluation score based on the chosen figure of merit for all training circuits.
Afterwards, the ML model can be \mbox{re-trained} in minutes.
This is enabled by persistently storing the compiled circuits and evaluation scores of previously conducted compilations as part of the training data generation process.
On the contrary, no additional effort is needed to exclude devices---either in case of a temporary downtime due to, e.g., calibration, or a permanent deprecation.
These devices can just be masked out without any further effort necessary.

\section{Conclusion}
\label{sec:conclusions}

Using quantum computing as a technology to \emph{just} solve a problem from any kind of application domain is challenging, since a suitable quantum device must be selected to solve a problem encoded as a quantum circuit and the circuit must be compiled accordingly.
Deciding which is the best device and how to best compile for a certain application---according to a \emph{figure of merit}---currently requires expert knowledge in quantum computing.
So far, especially \mbox{end-users} from application domains are often left unsupported and overwhelmed due to missing automation and tool support.
This situation is even worsened, since the result quality heavily fluctuates with the particular choice of options and often makes the difference between a useful result and a useless one due to too much noise.

In this work, we proposed the MQT~Predictor framework to automate the selection of suitable target devices and the subsequent compilation.
By that, \mbox{end-users} are freed from this task with the goal of preventing a scenario where quantum computers can only be utilized by quantum computing experts---hindering the overall adoption of the technology.
The proposed methodology allows learning on the basis of previously conducted compilations of quantum circuits for various devices.
For that, the problem is tackled from two angles:
Using \emph{Reinforcement Learning}, optimal compilation pass sequences are learned for all available devices---mixing and matching compilation passes from various sources and quantum SDKs.
Furthermore, using \emph{Supervised Machine Learning}, a prediction of a quantum device is made for a given quantum circuit according to customizable figures of merit such as, e.g., the \emph{expected fidelity} of the \mbox{to-be-compiled} quantum circuit.

The proposed framework has been implemented using more than $500$ quantum circuits from $2$ to $90$ qubits for seven devices from two qubit technologies and has shown to yield compiled circuits that are within the \mbox{top-3} in more than $98\%$ of cases while frequently outperforming any tested combination by up to $53\%$ when considering the expected fidelity as the figure of merit.
Furthermore, trained and evaluated for \emph{critical depth} as another figure of merit, the performance was within the \mbox{top-3} in $89\%$ of cases while frequently outperforming any tested combination by up to $400\%$.
The corresponding framework (which is part of the \emph{Munich Quantum Toolkit}~(MQT)) including its \mbox{pre-trained} models and classifiers is publicly available on GitHub (\url{https://github.com/cda-tum/mqt-predictor}) and as an \mbox{easy-to-use} \emph{Python} package (\url{https://pypi.org/p/mqt.predictor}).

These results clearly demonstrate that adapting technologies from the classical realm and applying them to quantum computing can significantly foster the progress of this new and promising technology.
Especially since quantum computers are mostly utilized by experts at the moment due to the challenges within the current workflow for realizing quantum application on real machines.
With this work, we hope to take a step forward in simplifying the utilization of quantum computers for \mbox{end-users} from application domains and the development of optimized compilers within the community.

\section*{Acknowledgments}
This work received funding from the European Research Council (ERC) under the European Union’s Horizon 2020 research and innovation program (grant agreement No. 101001318), was part of the Munich Quantum Valley, which is supported by the Bavarian state government with funds from the Hightech Agenda Bayern Plus, and has been supported by the BMWK on the basis of a decision by the German Bundestag through project QuaST, as well as by the BMK, BMDW, the State of Upper Austria in the frame of the COMET program, and the QuantumReady project within Quantum Austria (managed by the FFG).

\clearpage

\printbibliography

\clearpage
 \appendix
  
  \section{Appendices}
  \subsection{Device Distribution} \label{appendix:dev_dist}
  \begin{figure*}[t]
     \begin{subfigure}[b]{0.49\textwidth}
         \centering
         \includegraphics[width=\textwidth]{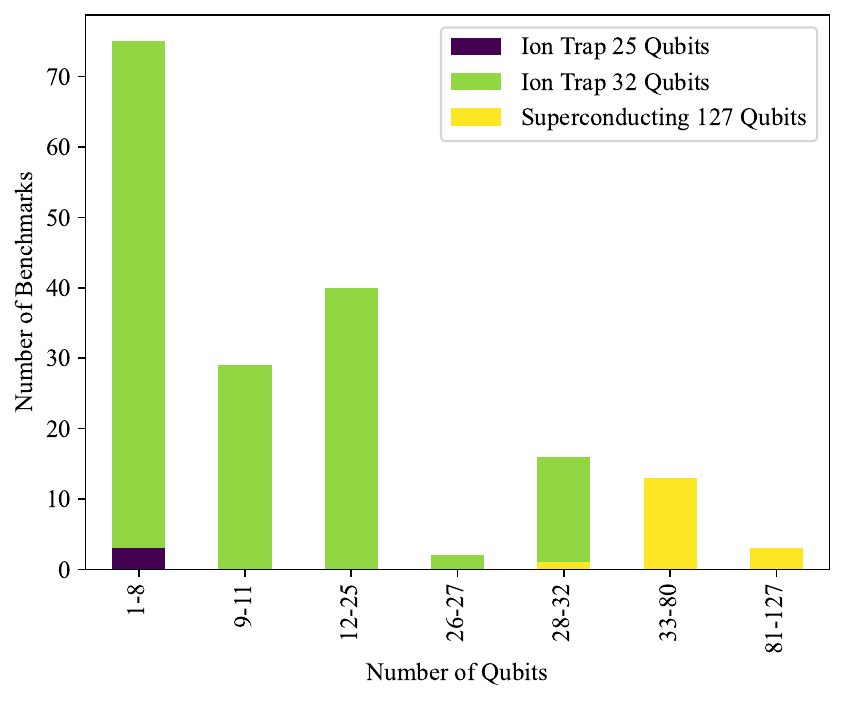}
         \caption{Expected fidelity.}
         \label{fig:device_distributions_exp}
     \end{subfigure}
     \hfill
     \begin{subfigure}[b]{0.49\textwidth}
         \centering
         \includegraphics[width=\textwidth]{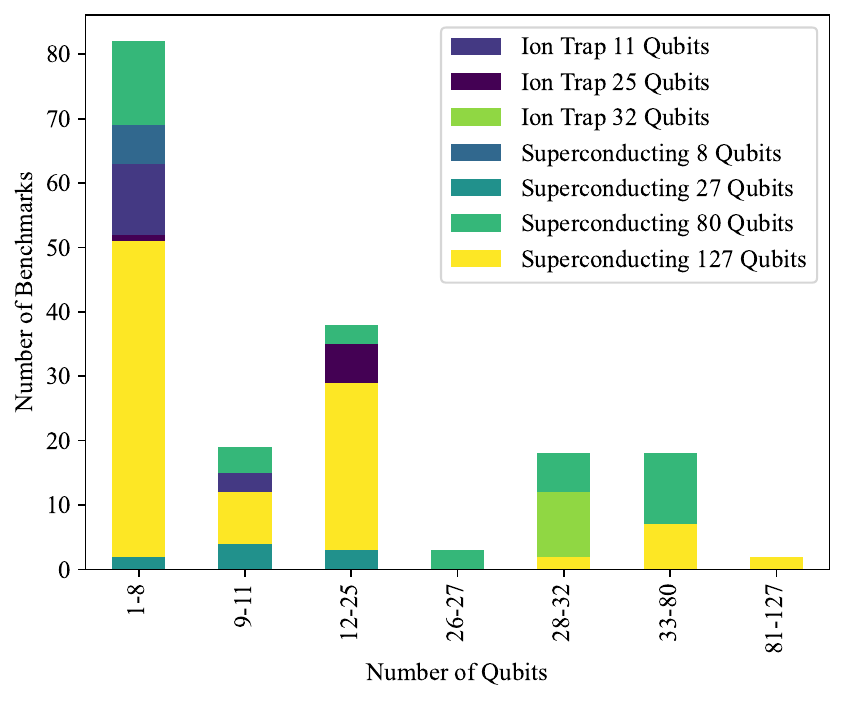}
         \caption{Critical depth.}
         \label{fig:device_distributions_crit}
     \end{subfigure}
\caption{Device distributions.}
\label{fig:device_distributions}
\end{figure*}

To better understand the problem of selecting the best device, the underlying ground truth data used to train the MQT~Predictor is further investigated.
To this end, the frequency with which each device is selected as the best performing one is investigated.
For that, each benchmark is compiled for all supported devices using the respectively trained RL compilers and the one leading to the best result is determined.
The results are visualized in \autoref{fig:device_distributions}.
Furthermore, the benchmarks are grouped by their number of qubits such that all benchmarks within one group can be executed on the same number of devices---leading to seven groups since currently seven devices are supported with a decreasing number of available devices for an increasing number of qubits.

When using the expected fidelity as the figure of merit, the results are rather distinct as visualized in \autoref{fig:device_distributions_exp}.
For most benchmarks, the ion trap device with $32$ qubits results in the best performance.
Therefore, the rule of thumb known by quantum computing experts---taking an ion trap device when possible---has been confirmed.
However, when all ion trap devices' capacities are exceeded, the largest superconducting device is best even when a smaller device would have been available as well.
This showcases that another quantum expert rule of thumb---taking always the smallest but fitting device---cannot be confirmed by our experimental data.

When using the critical depth as the figure of merit, the device distribution is significantly more diverse as visualized in \autoref{fig:device_distributions_crit}\footnote{The benchmarks used for the evaluation differ between both figures of merit and, therefore, the number of benchmarks in each group differs as well.}.
Here, all seven devices are chosen at least sometimes and there is no apparent correlation between the best performing device and the benchmark groups since most devices result in the best performance for more than just one group.
Consequently, no clear guidance can be derived from that---underlining the importance of software tools to guide this decision automatically.

  \clearpage 
  \subsection{Detailed Evaluation: Device Selection Using ML} 
  \label{appendix:ml}
	
  \begin{figure*}[t]
     \begin{subfigure}[b]{0.49\textwidth}
         \centering
         \includegraphics[width=\textwidth]{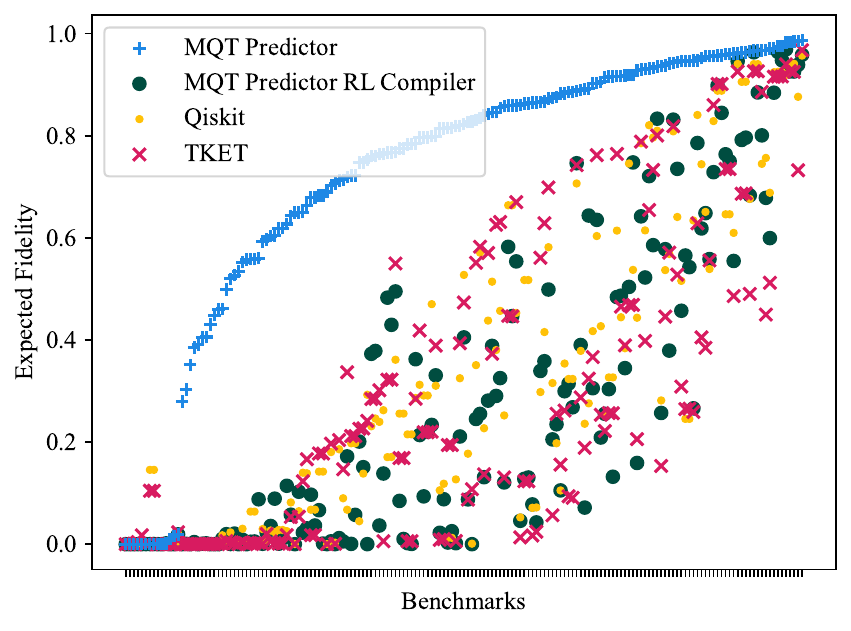}
         \caption{Expected fidelity.}
         \label{fig:detailed_ml_fid}
     \end{subfigure}
     \hfill
     \begin{subfigure}[b]{0.49\textwidth}
         \centering
         \includegraphics[width=\textwidth]{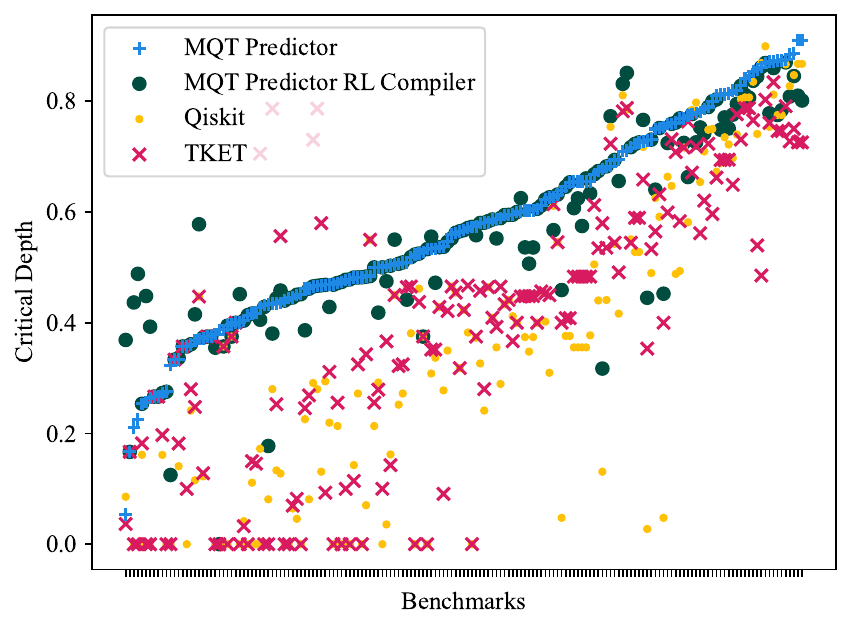}
         \caption{Critical depth.}
         \label{fig:detailed_ml_crit}
     \end{subfigure}
\caption{Isolated evaluation of the device selection using ML.}
\label{fig:detailed_ml_results}
\end{figure*}  
  
  The MQT~Predictor framework automatically selects the most promising device and compiles accordingly. 
  To assess the effect each of the two tasks has, they are evaluated in an isolated fashion---starting with the device selection and its respective results shown in \autoref{fig:detailed_ml_results}.
	To this end, the largest device (superconducting device with $127$ qubits) is used as the default device to fit all benchmarks.
	Then, the benchmarks are compiled for it using Qiskit, TKET, and the MQT~Predictor RL compiler and, similarly to \autoref{fig:fidelity_plot} and \autoref{fig:depth_plot}, the benchmarks are sorted by their MQT~Predictor score for better readability.
	The resulting performances are then compared against the full MQT~Predictor approach including the device selection.
	
For the expected fidelity, the results are visualized in \autoref{fig:detailed_ml_fid}---showing a distinct gap between the results of the full MQT~Predictor approach including the device selection compared to all three baselines using the default device.
This highlights the importance of selecting the most suitable device, since that can make the difference between a successful execution and obtaining completely random results.
Furthermore, it is interesting to see that there is no clear winner among Qiskit, TKET, and the MQT~Predictor RL compiler.

In the case of critical depth, the situation is different as visualized in \autoref{fig:detailed_ml_crit} since the full MQT~Predictor approach less frequently leads to the best result.
This suggests that the default device is a good choice for critical depth and the influence between it and the best one is smaller than it is the case for the expected fidelity.
Also, this matches with \autoref{fig:device_distributions_crit} since often the largest device is the most suitable in the ground truth data.
Furthermore, the MQT~Predictor RL compiler approach outperforms both Qiskit and TKET in most cases---showcasing, that both are not optimizing for critical depth but optimize for a different implicit underlying criterion closer to the expected fidelity.

  \clearpage 
  \subsection{Detailed Evaluation: Compilation Using RL} 
  \label{appendix:rl}

  \begin{figure*}[t]
     \begin{subfigure}[b]{0.49\textwidth}
         \centering
         \includegraphics[width=\textwidth]{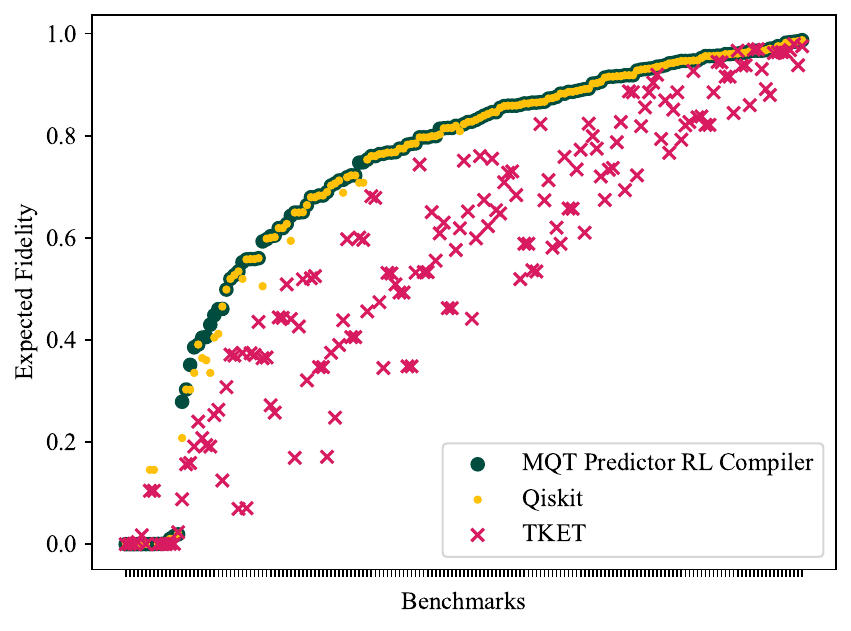}
         \caption{Expected fidelity.}
         \label{fig:detailed_rl_fid}
     \end{subfigure}
     \hfill
     \begin{subfigure}[b]{0.49\textwidth}
         \centering
         \includegraphics[width=\textwidth]{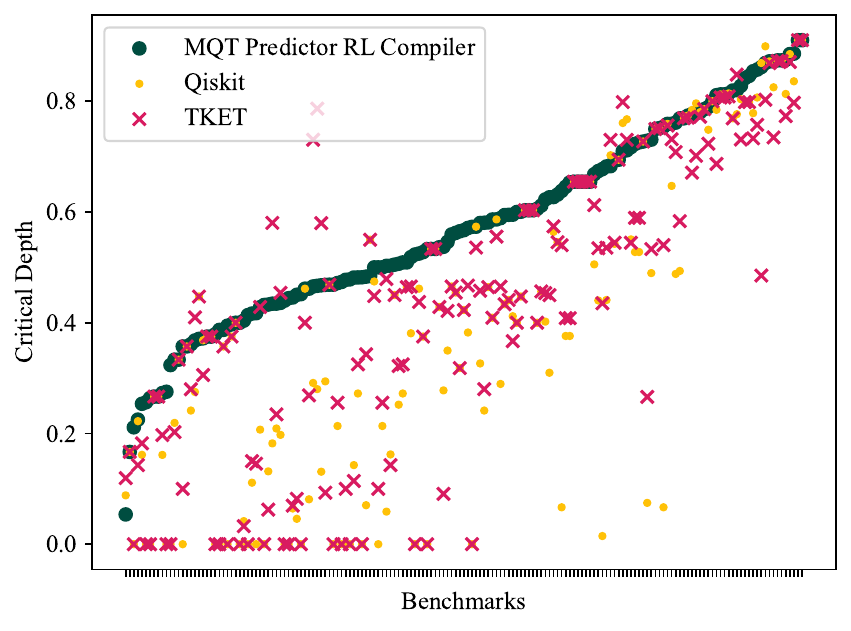}
         \caption{Critical depth.}
         \label{fig:detailed_rl_crit}
     \end{subfigure}
\caption{Isolated evaluation of the compilation using RL.}
\label{fig:detailed_rl_results}
\end{figure*}

To assess the compilation quality of the RL part of the MQT~Predictor framework, it is evaluated in a similar and isolated fashion as the device selection described in \aref{appendix:ml}.
To this end, the device is set to be the one predicted for each benchmark and the effect of the different compilers is evaluated in \autoref{fig:detailed_rl_results}.
For that, the same compilers as before are used: Qiskit, TKET, and the MQT~Predictor RL compiler---while, in this case, the latter corresponds to the full MQT~Predictor approach because the default device is the predicted one.
Therefore, this evaluation setup results in plots that show a subset of the data points plotted in \autoref{fig:fidelity_plot} respectively \autoref{fig:depth_plot} in which the compilers are evaluated for all seven devices while here only the predicted one is used.
The benchmarks are sorted by their MQT~Predictor RL compiler scores for better readability (which, in this case, corresponds to the scores of the full MQT~Predictor approach).

For the expected fidelity visualized in \autoref{fig:detailed_rl_fid}, the MQT~Predictor RL compiler outperforms TKET and Qiskit, although there is only a small performance gap to the latter since Qiskit performs better than TKET for the predicted device. 
Compared to \autoref{fig:detailed_ml_fid} where the default device was the largest superconducting one, this was not the case. 
Therefore, apparently the compilation quality significantly depends on the device---either by Qiskit being better for the predicted one (which is often an ion trap one), TKET being worse, or both.
Nevertheless, the plot shows the (small) improvement of the MQT~Predictor compilation---potentially growing when including more compilation passes from further providers.

In the case of the critical depth visualized in \autoref{fig:detailed_rl_crit}, the performance differences between Qiskit and TKET are not as distinct as before. 
Furthermore, the performance difference of the MQT~Predictor RL compiler compared to both of them is even larger---highlighting the impact it has.

\end{document}